\documentclass[utf8]{FrontiersinHarvard} 

\usepackage{url,hyperref,lineno,microtype,subcaption,upgreek,amsmath ,mathrsfs}
\usepackage[onehalfspacing]{setspace}

\def\keyFont{\fontsize{8}{11}\helveticabold }
\def\firstAuthorLast{Zhang {et~al.}} 
\def\Authors{Xiao-Li Zhang\,$^{1}$,  Yong-Feng Huang\,$^{2,3*}$ and Ze-Cheng Zou\,$^{2}$   }
\def\Address{$^{1}$Department of Physics, Nanjing University, Nanjing 210093, China \\
$^{2}$School of Astronomy and Space Science, Nanjing University, Nanjing 210023, China \\
$^{3}$Key Laboratory of Modern Astronomy and Astrophysics (Nanjing University), Ministry of Education, Nanjing 210023, China }
\def\corrAuthor{Yong-Feng Huang}

\def\corrEmail{hyf@nju.edu.cn}

\begin{document}
\onecolumn
\firstpage{1}

\title[Strange quark stars]{Recent progresses in strange quark stars}

\author[\firstAuthorLast ]{\Authors} 
\address{} 
\correspondance{} 

\extraAuth{}

\maketitle

\begin{abstract}

According to the hypothesis that strange quark matter may be the
true ground state of matter at extremely high densities, strange
quark stars should be stable and could exist in the Universe. It
is possible that pulsars may actually be strange stars, but not
neutron stars.  Here we present a short review on recent
progresses in the field of strange quark stars. First, three
popular phenomenological models widely used to describe strange
quark matter are introduced, with special attention being paid on
the corresponding equation of state in each model. Combining the
equation of state with the Tolman-Oppenheimer-Volkov equations,
the inner structure and mass-radius relation can be obtained for
the whole sequence of strange stars.
Tidal deformability
and oscillations (both radial and non-radial oscillations), which
are sensitive to the composition and the equations of state, are
then described. Hybrid stars as a special kind of quark stars are
discussed. Several other interesting aspects of strange stars are
also included.  For example, strong gravitational wave emissions
may be generated by strange stars through various mechanisms,
which may help identify strange stars via observations.
Especially, close-in strange quark planets with respect to their
hosts may provide a unique test for the existence of strange quark
objects. Fierce electromagnetic bursts could also be generated by
strange stars. The energy may come from the phase transition of
neutron stars to strange stars, or from the merger of binary
strange stars. The collapse of the strange star crust can also
release a huge amount of energy.  It is shown that strange quark
stars may be involved in short gamma-ray bursts and fast radio
bursts.

\section{}


\tiny
 \keyFont{ \section{Keywords:} stars: neutron, dense matter, equation of state,
                 gravitational waves, gamma-ray bursts, fast radio bursts}
\end{abstract}

\section{Introduction}

Strange quark matter, which is a mixture consisting of almost equal numbers of
deconfined up, down, and strange quarks, may be true ground state of dense
matter \citep{Witten_1984_PhysRevD,Farhi&Jaffe_1984_PhysRevD}.  If such a strange quark matter hypothesis
is correct, then the observed pulsars may actually be strange quark stars (also
shortened as strange stars). Strange quark stars (SQS), which involve extraordinarily high
densities, intense gravitational fields and strong electromagnetic
fields \citep{Alcock_1986_ApJ,Colpi_1992_ApJ}, provide ideal natural laboratories
for exploring physics under extreme astrophysical conditions. However, the nature
of the strongly interacting matter under such extreme densities is still quite
unclear, leading to large uncertainties in the internal structure of strange stars.
A lot of efforts have been devoted to the theoretical and observational aspects
of strange stars, but many issues still remain unsolved in the field.

Strange quark matter is inherently self-bound by the strong interactions of quarks.
As a result, compact stars composed of strange quark matter could be bare strange
stars whose density reduces to zero abruptly at the surface. However, a strange
star can also have a thin crust composed of normal nuclear matter \citep{Glendenning_1992_ApJ}.
Because the maximum density of the crust is five times lower than the neutron drip
density \citep{Huang_1997_A&A}, the light crust has an  almost negligible effect on
the internal structure of strange stars \citep{Zdunik_2002_A&A}. However, it could
significantly change the characteristics of electromagnetic emissions from such compact
stars. Also, collapse of the crust could lead to some kinds of electromagnetic bursts or
even emission of gravitational waves.

Interestingly, according to the strange quark matter hypothesis, strange quark dwarfs
and even strange quark planets could also exist. They may be produced due to the
contamination of white dwarfs/planets by strange nuggets in the Universe. While the stability
of these low-mass strange objects is still highly debated
\citep{Glendenning_1995_ApJ,Fraga_2001_PRD,Vartanyan_2012_Ap,Vartanyan_2014_JPhCS,Alford_2017_ApJ,DiClemente_2023_A&A,Goncalves_2023_EPJA},
they could provide valuable opportunities for identifying strange stars
due to their significant difference from normal matter white dwarfs and planets
\citep{Huang_2017_ApJ,Kuerban_2020_ApJ,Wang_2021_PhRvD,Kurban_2022_PhLB}.

The study of strange stars is an active and rapidly developing
field. In the past few decades, various observational aspects of
strange stars have been studied. In this review, we are going to
present a brief description on some recent progresses concerning
strange stars. Hybrid stars, in which quark matter and
hadronic matter may coexist, are also discussed.  The structure
of our paper is organized as follows. The properties and internal
structure of strange stars are introduced in Section \ref{InternalStructure}. Especially, several popular phenomenological models
describing the strong interaction among quarks are presented.
Tidal deformability and oscillations, which are closely related to the internal structure, are also introduced.
As a special kind of quark stars, hybrid stars are discussed in Section
\ref{HybridStars}, paying special attention on the transition between different phases.
In Section \ref{GW}, gravitational wave (GW) emissions from
various strange quark objects are discussed. Possible connection
between some violent electromagnetic bursts (e.g. gamma-ray bursts
and fast radio bursts) and strange stars are introduced in Section
\ref{ElectromagneticBursts}. Finally, Section
\ref{ConcludingRemarks} presents the conclusions and some
further discussion.

\section{Internal structure of strange stars}
\label{InternalStructure}

While thousands of pulsars have been observed, the internal composition and structure of
them are still controversial. This enigma is closely connected with the interaction and
properties of matter under extreme densities and temperatures, thus is an important
issue involving fundamental physics. Theoretically, pulsars could be neutron stars or
quark stars. In the quark star case, they could either be three-flavor (u, d, s) strange
stars, or could even be two-flavor (u, d) quark stars. The existence of the
so called hybrid stars is also suggested, which usually include a quark matter core encompassed
by normal nuclear matter \citep{1965Ap......1..251I,2020NatPh..16..907A,2021Univ....7..267M}.

Equation of state (EOS), which gives the relation between pressure and energy density,
is an important factor that determines the structure and overall properties of compact
stars. EOS is generally dependent on the composition and temperature of the dense matter.
In principle, EOS could be derived by considering the strong interaction of microscopic
particles that constitute the dense matter. However, due to the complexity of strong
interaction and our poor understanding on it, an accurate derivation of the EOS is still
impossible. Various models have been proposed to describe the strong interaction of quarks.
In this section, we will introduce several widely used quark interaction models. The internal
structure of compact stars based on these models will also be addressed. Note that some
further complicated ingredients such as the magnetic field and the spin of the star should
also be included when they play a non-negligible role in some extreme cases.

\subsection{MIT bag model}

The MIT bag model, initially proposed in the 1970s by \cite{Chodos_1974_1_PhRvD,Chodos_1974_PhRvD},
is a phenomenological theoretical description aiming at explaining the structure of hadrons. In this
framework, the finite space containing hadrons is regarded as a ``bag''. Hadrons in the bag are
composed of free quarks, including up, down, and strange quarks in case of strange stars.
Such a confinement is not a dynamical outcome of any underlying theory but rather a feature
imposed by hand, achieved through the imposition of particular boundary
conditions \citep{Buballa_2005_PhR}.

The general form of the EOS of this model is \citep{Shafeeque_2023_JApA, Lohakare_2023_MNRAS}
\begin{eqnarray}
    P = k\big( \epsilon c^2 - 4B\big),
    \label{funcMIT}
\end{eqnarray}
where $P$ denotes the pressure within the bag, $\epsilon$ represents the energy density,
and $c$ is the velocity of light. The parameter $k$  is a constant that depends upon
the mass of strange quarks  ($m_s$) and the quantum chromodynamics (QCD) coupling constant. Specifically,
for $m_s=0$, $k=1/3$. $B$ signifies the pressure of the bag, corresponding to the
energy density when the bag is in a vacuum state. It is a free parameter that could
be determined by considering the hadron spectra or other experimental data.

The bag constant is also equivalent to the critical pressure of deconfinement, resulting
in a pressure differential across the surface of the bag. The characteristics of quark
confinement can be described qualitatively by the bag constant. However, it is too
simple to describe asymptotic freedom at increasing energy scales, a crucial property
associated with QCD. Anyway, the model has only one
free parameter ($B$), making it the simplest theory to compute the EOS of strange
quark matter.

\subsection{NJL model}

The Nambu-Jona-Lasinio (NJL) model is initially proposed
by \cite{Nambu_1961_PhRv,Nambu_1961_1_PhRv} to describe the interaction between
nucleons. It was extended by \cite{Eguchi_1974_PhRvD} to include up and down
quarks. \cite{Kikkawa_1976_PThPh} further developed the theory to encompass
three flavors of quarks, including up, down, and strange quarks. In this review, we utilize
the three-flavor NJL model to describe quark matter inside strange stars.
The model exhibits spontaneous breaking of chiral symmetry, which is essential for
understanding the large nucleon mass and the dynamic generation of fermion masses.
Additionally, the model is noteworthy for its solvability, allowing for simple
analytical results obtained in certain limiting cases.

The Lagrangian of three-flavor NJL model is generally expressed as
\begin{eqnarray}
    \mathcal{L} = \bar{\psi}\left(i \gamma^\mu \partial_\mu - m\right) \psi + \mathcal{L}^{(4)} + \mathcal{L}^{(6)} ,
\end{eqnarray}
where $\psi = (\psi_u, \psi_d, \psi_s)^\mathrm{T}$ represents the quark
field, $\gamma^\mu$ denotes the Dirac matrix, $m= diag(m_u, m_d, m_s)$
is the fluid quark mass matrix, and $\partial_\mu$ is the partial differential
operator in four-dimensional spacetime. $\mathcal{L}^{(4)}$
and $\mathcal{L}^{(6)}$ correspond to four-fermion interaction and
six-fermion interaction terms, respectively, representing two-body and
three-body interactions. The isospin symmetry indicates that $m_u = m_d$,
but $m_s$ differs from both $m_u$ and $m_d$, thereby manifesting the
$SU(3)$ symmetry breaking in the three-flavor NJL model.

Since quark confinement is not directly reflected in the NJL model, it is often
used in conjunction with the bag constant ($B$) to derive the EOS. Comparing
with the MIT bag model, the confinement is a dynamical outcome in the NJL
model rather than being imposed by hand. However, similar to the MIT bag
model, the NJL model also fails to explain the characteristics of asymptotic freedom.
Additionally, in this model, the interaction between quarks is regarded as
point-to-point interaction without the inclusion of gluons. Consequently, it is
not a renormalizable field theory, and a regularization scheme must be specified
to address the improper integrals that arise \citep{Klevansky_1992_RvMP}. For
more details, readers can refer to the review articles
by \cite{Klevansky_1992_RvMP} and \cite{Buballa_2005_PhR}.

\subsection{Quasi-particle model}

The quasi-particle model is another phenomenological approach for strange quark matter.
\cite{Peshier_1994_PhLB} and \cite{Gorenstein_1995_PhRvD} initially employed this model
to describe the quark-gluon plasma with strong interactions. Here we present a short
introduction to the model. First, the pressure at zero temperature and finite chemical
potential can be expressed as a model-independent
formula \citep{He_2007_JPhG,Zong_2008_PhRvD,Zong_2008_IJMPA},
\begin{eqnarray}
    P(\mu) = P(\mu) \rvert_{\mu=0} + \int_{0}^{\mu}\mathrm{d}\mu'\,\rho\big(\mu'\big),
\end{eqnarray}
where $\mu$ denotes the chemical potential, $P(\mu) \rvert_{\mu=0}$ signifies the
pressure at zero chemical potential, and $\rho(\mu)$ is the number density of quarks.

To calculate the pressure, the primary challenge lies in computing the number density
of quarks, which relies on the quark propagator. However, calculating the quark
propagator directly from the first principles of QCD is impractical. Therefore, we have
to employ approximations and then use the quasi-particle model. In this model,
particles are treated as an ideal gas composed of non-interaction quasi-particles,
with their masses depending on the temperature and density. This model simplifies
the calculation of particle interactions, making computations more tractable.
In this way, the EOS is derived in the framework of the quasi-particle model as \citep{Zhao_2010_MPLA}
\begin{eqnarray}
    P(\mu) = P(\mu) \rvert_{\mu=0} + \frac{3}{\pi^2} \int_{0}^{\mu} \mathrm{d}\mu'\,\theta\Big(\mu' - m\big(\mu'\big)\Big)\Big(\big(\mu'\big)^2 - m\big(\mu'\big)^2\Big)^{3/2} ,
\end{eqnarray}
where $m(\mu)$ is the effective mass and $\theta$ is the Heaviside step function.
Here, three colors and three flavors have been considered for quarks.

In view of quark confinement, the energy density in the vacuum is
lower than that of free quarks. Consequently, the vacuum pressure
at zero chemical potential ( $P(\mu) \rvert_{\mu=0}$ ) is
negative. Using the MIT bag model for reference, we can set
$P(\mu) \rvert_{\mu=0} = -B$.
In this framework,
quarks are treated as Fermi gas at high chemical potentials, which
to some extent reflects the asymptotic freedom property of QCD. It
also matches well with the results of Lattice QCD
\citep{Peshier_1999_PRC, Peshier_2002_PhRvD,Rebhan_2003_PhRvD}.

\subsection{Tolman-Oppenheimer-Volkoff equation}

Due to the extremely high density of strange stars
accompanied by strong gravity, the effects of spacetime curvature
cannot be ignored. Consequently, the structure of strange stars
has to be studied in the context of General Relativity. The
Tolman-Oppenheimer-Volkoff (TOV) equation
\citep{Tolman_1939_PhRev,Oppenheimer_1939_PhRv}, should be
employed to infer the structure of such compact objects, which
assumes that the interior of the star is composed of spherically
symmetric ideal fluid. Using spherical coordinates of $ (x^0, x^1,
x^2, x^3) = (t, r, \theta, \phi) $, the generalized Schwarzschild
metric is \citep{Oppenheimer_1939_PhRv}
\begin{eqnarray}
    ds^2 = -e^{\sigma(r)} dt^2 + e^{\eta(r)} dr^2 + r^2 d\theta^2 + r^2 sin^2\theta d\phi^2 ,
    \label{ds2}
\end{eqnarray}
where $\sigma(r)$ and $\eta(r)$ are functions of radial
coordinates. Comparing it with the spacetime interval, $ds^2 =
g_{ab}dx^a dx^b $, the nonzero covariant components of the
metric tensor $g_{ab}$ can be obtained as
\begin{eqnarray}
    g_{tt} = -e^{\sigma(r)}, \indent
    g_{rr} = e^{\eta(r)}, \indent
    g_{\theta\theta} = r^2, \indent
    g_{\phi\phi} = r^2 sin\theta^2 .
    \label{MetricTensor}
\end{eqnarray} 

The energy-momentum tensor of such a ideal fluid is given
by 
\begin{eqnarray}
    T^{ab} = (p + \epsilon)u^{a}u^{b} + p g^{ab} ,
    \label{Energy-momentumTensor}
\end{eqnarray}
where $p$ is the pressure and $\epsilon$ is the energy density.
For a static spherically symmetric star, the pressure and energy
density are functions of the r-coordinate, i.e. $p(r)$ and
$\epsilon(r)$. $u^{a}$ and $u^{b}$ are four-dimension
velocities.
Equation (\ref{Energy-momentumTensor}) can be rewritten as
\begin{eqnarray}
    T^{a \phantom{j}}_{\phantom{k}b} = (p + \epsilon)u^{a}u_{b} + p \delta^{a \phantom{j}}_{\phantom{k}b} .
    \label{Energy-momentumTensor2}
\end{eqnarray}
Since the star is static, there are no spatial components for the
four velocities, i.e. $ u^i = 0 (i=1,2,3), u^0 = u^t =
1/\sqrt{-g_{tt}}$. Considering the normalization condition of
$u^a u_b =-1 $, we have \citep{Tolman_1934_rtc_book}
\begin{alignat}{2}
    & T^{0 \phantom{j}}_{\phantom{k}0} = -\epsilon , \\
    & T^{a \phantom{j}}_{\phantom{k}a} = p \quad (a =
    1,2,3).
\end{alignat} 

The Einstein's field equation is
\begin{eqnarray}
    R_{a b}- \frac{1}{2} R g_{a b} + \Lambda g_{a b} = \frac{8\pi G}{c^4} T_{a b} ,
    \label{EinsteinFieldEquation}
\end{eqnarray}
where $R_{a b}$ is the Ricci curvature tensor, $R$ is the
scalar curvature, $\Lambda$ is the cosmological constant, $G$ is
Newton's gravitational constant and $c$ is the speed of light. For
simplicity, we take $c=1$ hereinafter. The cosmological constant
$\Lambda$ is negligible for compact stars. 

The Ricci tensor is expressed as \citep{Glendenning_1996_book}
\begin{eqnarray}
    R_{ab} = \Gamma^\alpha_{a\alpha,b} -  \Gamma^\alpha_{ab,\alpha} + \Gamma^\alpha_{a\beta}\Gamma^\beta_{b\alpha} - \Gamma^\alpha_{ab}\Gamma^\beta_{\alpha\beta} ,
    \label{RicciTensor}
\end{eqnarray}
where the Christoffel symbol is defined as
\begin{eqnarray}
    \Gamma^\lambda_{ab} = \frac{1}{2}g^{\lambda\kappa} (g_{\kappa a,b} + g_{\kappa b,a} - g_{ab,\kappa}) .
\end{eqnarray}
Using the metric tensor of Equation (\ref{MetricTensor}), we can
obtain the non-zero Christoffel symbols as \citep{Glendenning_1996_book}
\begin{alignat}{4}
    & \Gamma^1_{00} = \frac{1}{2}\sigma'e^{\sigma-\lambda} ,
    \indent\indent \Gamma^0_{10} = \frac{1}{2}\sigma' , \\
    & \Gamma^1_{11} = \frac{1}{2}\eta' ,
    \indent\indent\indent\indent\indent \Gamma^2_{12} = \Gamma^3_{13} = \frac{1}{r} , \\
    & \Gamma^1_{22} = -re^{-\eta} ,
    \indent\indent\indent\;\Gamma^3_{23} = cot\theta , \\
    & \Gamma^1_{33} = -r sin^2 \theta e^{-\eta} ,
    \indent \Gamma^2_{33} = -sin\theta cos\theta.
\end{alignat}
Here the primes denote differentiation with respect to the
r-coordinate. Then the nonzero components of the Ricci tensor in
Equation (\ref{RicciTensor}) is derived as \citep{Glendenning_1996_book}
\begin{alignat}{4}
    & R_{00} = e^{\sigma-\eta} (-\frac{\sigma''}{2}-\frac{\sigma'}{r}+\frac{\sigma'}{4}(\eta'-\sigma')) ,\\
    & R_{11} = \frac{\sigma''}{2}-\frac{\eta'}{r}+\frac{\sigma'}{4}(\sigma'-\eta') ,\\
    & R_{22} = e^{-\eta}(1 +\frac{r}{2}(\sigma'-\eta'))-1 ,\\
    & R_{33} = R_{22}sin^2\theta .
\end{alignat}
Furthermore, we have
\begin{eqnarray}
    R^{a \phantom{j}}_{\phantom{k}b} = g^{aa}R_{ab} .
\end{eqnarray}
The scalar curvature is obtained from the trace of the Ricci
tensor,
\begin{eqnarray}
    R = g^{ab} R_{ab} = R^{a\phantom{j}}_{\phantom{k}a} .
\end{eqnarray} 

The Einstein's field equation (\ref{EinsteinFieldEquation}) can be rewritten as, 
\begin{eqnarray}
    R^{a \phantom{j}}_{\phantom{k}b} - \frac{1}{2}R \delta^{a \phantom{j}}_{\phantom{k}b} = 8\pi G T^{a \phantom{j}}_{\phantom{k}b} .
    \label{EinsteinFieldEquation2}
\end{eqnarray}
Substituting $R^{a \phantom{j}}_{\phantom{k}b}$, $R$ and
$T^{a \phantom{j}}_{\phantom{k}b}$ into the equation, we have
\begin{alignat}{4}
    & R^{0 \phantom{j}}_{\phantom{k}0}-\frac{1}{2}R \delta^{0 \phantom{j}}_{\phantom{k}0}= e^{-\eta}(\frac{1}{r^2}-\frac{\eta'}{r}) -\frac{1}{r^2} = -8\pi G \epsilon , \\
    & R^{1 \phantom{j}}_{\phantom{k}1}-\frac{1}{2}R \delta^{1 \phantom{j}}_{\phantom{k}1}= e^{-\eta}(\frac{1}{r^2}+\frac{\sigma'}{r}) -\frac{1}{r^2} = 8\pi G p ,   \\
    & R^{2 \phantom{j}}_{\phantom{k}2}-\frac{1}{2}R \delta^{2 \phantom{j}}_{\phantom{k}2}= e^{-\eta}(\frac{\sigma''}{2}-\frac{\sigma'\eta'}{4} +\frac{\sigma'^2}{4}+\frac{\sigma'}{2r}-\frac{\eta'}{2r} ) = 8\pi G p ,  \\
    & R^{3 \phantom{j}}_{\phantom{k}3}-\frac{1}{2}R \delta^{3 \phantom{j}}_{\phantom{k}3}= e^{-\eta}(\frac{\sigma''}{2}-\frac{\sigma'\eta'}{4} +\frac{\sigma'^2}{4}+\frac{\sigma'}{2r}-\frac{\eta'}{2r} ) = 8\pi G p .
\end{alignat}
Note that the last two equations are identical, hence there are
only three independent equations. 

The boundary conditions can be taken as the Schwarzschild
metric vacuum solution, which gives \citep{Tolman_1939_PhRev}
\begin{eqnarray}
    -e^{\sigma(r)}= 1-\frac{2GM}{r} ,\;
    e^{\eta(r)}= -({1-\frac{2GM}{r}})^{-1} ,
\end{eqnarray}
where $M$ is the mass of the whole star. The TOV equation can then
be derived as \citep{Oppenheimer_1939_PhRv}
\begin{align}
    \frac{\mathrm{d}p(r)}{\mathrm{d}r} = -\frac{G(\epsilon +p)(m+4\pi r^3 p)}{r(r-2Gm)} .
    \label{TOV}
\end{align}
Here the mass included inside a sphere of radius $r$ is
\begin{eqnarray}
    m(r)=4\pi\int_{0}^{r}\epsilon(x)x^2dx ,
\end{eqnarray}
which can be equivalently expressed in the differential form of
\begin{eqnarray}
    \frac{\mathrm{d}m(r)}{\mathrm{d}r} = 4\pi r^2 \epsilon .
    \label{TOVm}
\end{eqnarray} 
Given the pressure and energy density at the center of the
star, i.e. $p(0)$ or $\epsilon(0)$, we can calculate the mass and
radius of a compact star by solving Equations (\ref{TOV}) and
(\ref{TOVm}). 

\begin{figure}[htbp]
    \includegraphics[width=.49\textwidth]{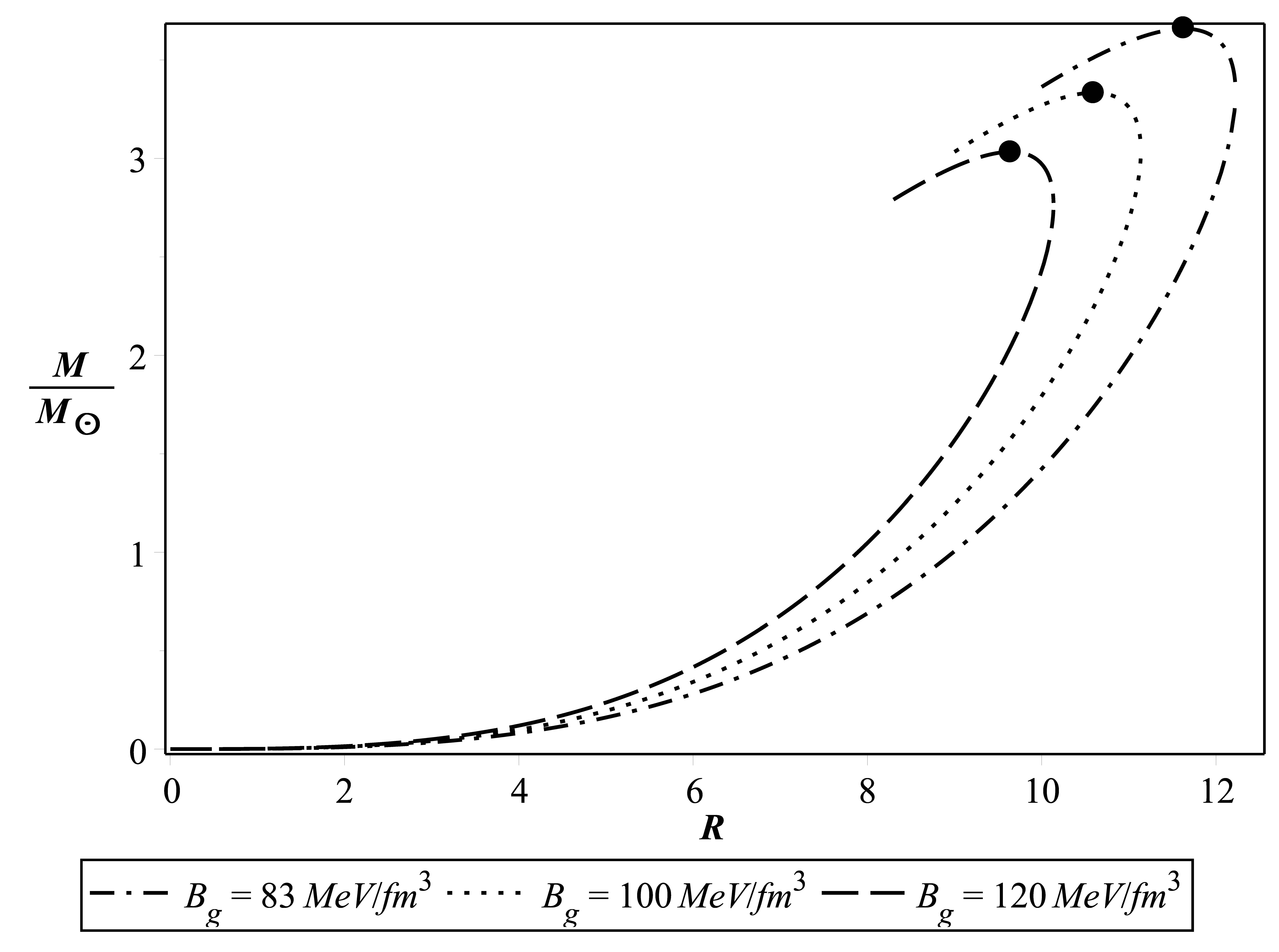}
    \includegraphics[width=.49\textwidth]{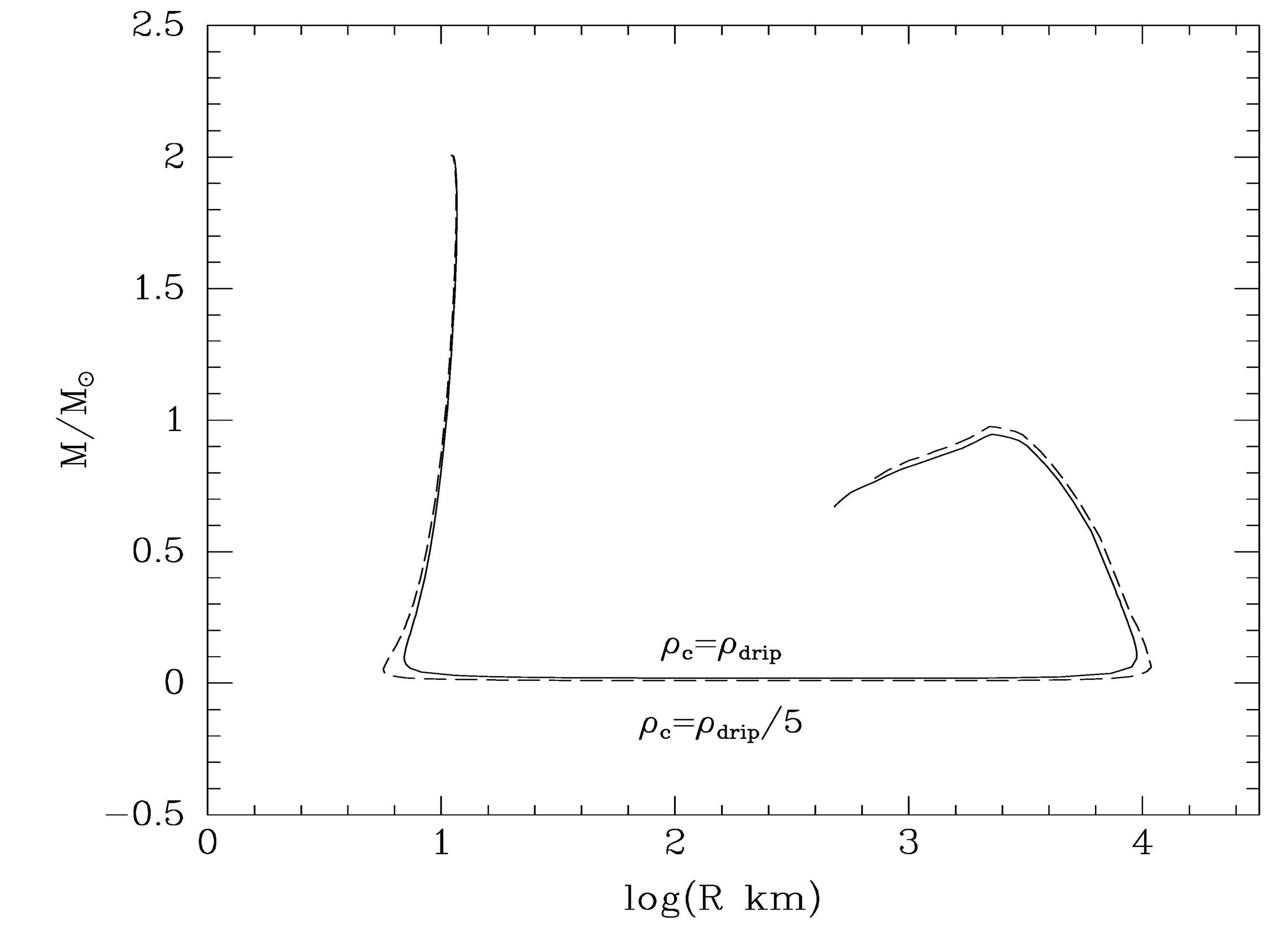}
    \centering
    \caption{Mass (normalized in solar masses) versus radius (km) for strange stars. \textit{Left panel:} bare strange stars.
                      Three different values are taken for the bag constant, i.e.  83, 100 and
                      120 MeV/fm$^3$. The filled circle on each curve represents the maximum-mass
                       star \citep{Deb_2017_AnPhy}. \textit{Right panel:} crusted strange stars.
                       The solid and dashed lines represent the bottom density of neutron drip
                       density ($\rho_\mathrm{drip}$) and $\rho_\mathrm{drip}/5$,
                       respectively \citep{Huang_1997_A&A}.}
    \label{Deb_2017_AnPhy}
\end{figure}

Typical mass-radius curves of bare strange stars derived by using the MIT bag model
are shown in the left panel of Figure \ref{Deb_2017_AnPhy}  \citep{Deb_2017_AnPhy}.
On the other hand, the mass-radius curves of crusted strange stars are shown in the right
panel of Figure \ref{Deb_2017_AnPhy}  \citep{Huang_1997_A&A}. This mass-radius relationship
help us understand the maximum and minimum limits of mass, the upper limit of radius, the
density distribution and even the stability condition of strange stars. We can also compare the
theoretically predicted parameters with observational data to test the model and gain
insights into the nature of matter under extreme conditions.

\subsection{Tidal deformability and Love numbers }

In the framework of General Relativity, an external tidal
field perturbs the spacetime geometry around a star, leading to
changes in the metric coefficients. These changes can be
analytically expressed in the asymptotic region far from the star.
 For a spherically symmetric static star with a mass of
$m$, the metric coefficient $g_{tt}$ in the local rest frame
(which is asymptotically Cartesian and mass-centered coordinates)
at a distance of $r$ can be given by solving the Einstein's field
equation, which is \citep{Thorne_1998_PhRvD}
\begin{eqnarray}
    -\frac{1+g_{tt}}{2} = -\frac{m}{r}-\frac{(3n^i n^j -\delta^{ij})Q_{ij}}{2r^3} +\mathscr{O}(r^{-4}) +\frac{E_{ij}}{2}r^2n^in^j+ +\mathscr{O}(r^{3}) ,
    \label{gtt}
\end{eqnarray}
where $n^i=x^i/r$. This expression is in units of $G=c=1$. The
first term on the right hand is a general relativistic correction
for the classical Newtonian gravitational potential. The symmetric
and traceless tensors $Q_{ij}$ and $E_{ij}$ represent the
quadrupole moment and the external quadrupolar tidal field,
respectively.   

The distortion of an object caused by external
gravitational forces can be described by a linear function between
the quadrupole moment $Q_{ij}$ and the external quadrupolar tidal
field $E_{ij}$ \citep{Hinderer_2008_apj,Flanagan_2008_PhRvD},
\begin{eqnarray}
    Q_{ij} = -\lambda E_{ij} ,
\end{eqnarray}
where the coefficient $\lambda$ is the so called tidal
deformability. It represents the extent to which the star deforms
under the influence of a specific tidal force. A commonly used
dimensionless tidal deformation parameter is then defined as,
\begin{eqnarray}
    \Lambda = \frac{\lambda}{m^5} .
\end{eqnarray} 

Another useful dimensionless tidal Love number ($k_2$)
connected with the coefficient $\lambda$ as
\begin{eqnarray}
    k_2 = \frac{3}{2} \lambda R^{-5} ,
\end{eqnarray}
where $R$ is the radius of the star. $k_2$ is called the Love
number, which represents the second spherical harmonic function
(for the angular quantum number $l=2$) used in calculating the
metric in Equation (\ref{gtt}). It can be calculated as
\cite{Hinderer_2008_apj}
\begin{align}
    k_2 &= \frac{8C^5}{5} (1-2C)^2 [2+2C(y-1)-y] \nonumber \\
        & \times \{2C [6-3y+3C(5y-8) ] \nonumber \\
        & + 4C^3 [13-11y+C(3y-2)+2C^2(1+y)] \nonumber \\
        & + 3(1-2C)^2 [2-y+2C(y-1)]\ln(1-2C)\}^{-1} ,
\end{align}
where $C=m/R$ is the compactness of the star. The function
$y=y(r)$ satisfies the differential equation of
\citep{Postnikov_2010_PRD,Tak_2020_PRD}
\begin{align}
    y(r) \frac{dy(r)}{dr} &+ y(r)^2 + Q(r)r^2 \nonumber \\
    & + y(r) e^{\eta(r)} [1+4\pi r^2 (p(r)-\epsilon(r))] = 0 ,
\end{align}
where $p(r)$ and $\epsilon(r)$ are the pressure and energy density, respectively. $Q(r)$ is expressed as
\begin{align}
    Q(r) &= 4\pi e^{\eta(r)} (5\epsilon(r)+ 9p(r) +\frac{p(r)+\epsilon(r)}{c_s^2(r)} ) \nonumber \\
        & - 6\frac{e^{\eta(r)}}{r^2} - (\sigma'(r))^2 ,
\end{align}
where $c_s(r)= \sqrt{dp/d\epsilon}$ is the sound speed.
$\sigma(r)$ and $\eta(r)$ are metric functions defined in Equation
(\ref{ds2}). Taking the boundary condition as $y(0)=2$
\citep{Damour_2009_PRD}, the tidal Love number $k_2$ can be
conveniently calculated by solving the TOV equation. 

The tidal Love number $k_2$ is strongly dependent on the
EOS of a star. Different EOSs of neutron stars and strange quark
stars result in different $k_2$. Thus the tidal deformability
could be a useful probe that could help distinguish between
neutron stars and strange stars \citep{Postnikov_2010_PRD}. 

\subsection{Oscillations and quasi-normal modes}

Oscillations are closely relevant to the stability of
stars and are sensitive to the equation of state and the
composition. In this aspect, quasi-normal modes are usually
discussed instead of normal modes, because the stars are
practically in a system with energy losses due to gravitational
radiation and other dissipative effects. The frequency of a
quasi-normal mode is usually expressed as a complex number, in
which the real part represents the actual oscillation frequency
and the imaginary part indicates the decay rate.

Quasi-normal modes include two parts, the radial
oscillation and non-radial oscillation. Radial oscillation refers
to the oscillation of the compact star in the radial direction.
Non-radial oscillation refers to the oscillation in a non-radial
direction, which is usually described by spherical harmonics.
Quasi-normal modes of non-radial oscillation are more complex and
involve different mode types, such as f-mode, p-mode, and g-mode.
Studying these quasi-normal modes can reveal oscillatory behaviors
of compact stars under gravitational wave emissions and other
dissipative effects, and help probe their internal structure.

\subsubsection{radial oscillations}

Radial oscillations of compact stars are investigated
firstly by \cite{Chandrasekhar_1964_PhRvL,Chandrasekhar_1964_ApJ}.
Usually adiabatic oscillations are considered: the whole star
oscillates like a retractable spring to expand and shrink
periodically. Assuming a spherical symmetry and considering
small-amplitude radial oscillations, we can introduce a
time-dependent radial displacement $\delta r(r,t)$ of a fluid
element at $r$, which is expressed as
\begin{eqnarray}
    \delta r(r,t) = X(r)e^{i \omega t} ,
\end{eqnarray}
where $X(r)$ represents the oscillation amplitude and $\omega$ is
the oscillation frequency. 

Under small perturbations, the contributions from
nonlinear terms can be ignored. The differential equation of the
radial displacement ($X(r)$) was derived by
\cite{Chandrasekhar_1964_ApJ} as 
\begin{align}
    Y\frac{d^2 X}{d^2 r}
    & + (\frac{dY}{dr}-Z+ 4\pi r \gamma p e^\eta -\frac{1}{2}\frac{d\sigma}{dr})\frac{dX}{dr} \nonumber \\
    & +[\frac{1}{2}(\frac{d\sigma}{dr})^2 +\frac{2m}{r^3}e^\eta -\frac{dZ}{dr} -4\pi(\epsilon +p)Zre^\eta +\omega^2 e^{\eta-\sigma}]X = 0 ,
\end{align}
where
\begin{eqnarray}
    Y(r) = \gamma p/ (\epsilon +p) ,
\end{eqnarray}
\begin{eqnarray}
    Z(r) = Y (-\frac{2}{r} +\frac{1}{2}\frac{d\sigma}{dr}) .
\end{eqnarray}
Here $\sigma(r)$ and $\eta(r)$ are again the functions defined in
Equation (\ref{ds2}). $\gamma$ is the adiabatic index determined
by the equation of state,
\begin{eqnarray}
    \gamma = \frac{\epsilon +p}{p} \frac{dp}{d\epsilon}.
\end{eqnarray}
Using the adiabatic index, $Y$ can be expressed as
\begin{align}
    Y(r) & = \frac{dp}{d\epsilon} \nonumber \\
    & \equiv c_s ^2 ,
\end{align}
where $c_s$ is the sound speed. 

To calculate $X(r)$, two boundary conditions must be
specified. First, the fluid at the center of the star is assumed
to remain at rest, i.e.
\begin{eqnarray}
    X(0) = 0 .
\end{eqnarray}
Second, the pressure perturbation vanishes at the stellar
surface, which means the Lagrangian variation of the pressure
should also vanish, i.e. 
\begin{eqnarray}
    \Delta p =-e^{\sigma/2} r^{-2} \gamma p \frac{d}{dr} (r^2 e^{-\sigma/2} X) =0 .
\end{eqnarray} 
Under these conditions, the eigenvalue of $\omega^2$ and
the corresponding radial eigenfunction of $X(r)$ can be solved.
The radial oscillation frequencies are directly linked to the mean
density and elastic properties, providing information about the
internal pressure and density distribution inside the star
\citep{Benvenuto_1991_MNRAS,Vaeth_1992_A&A,Jimenez_2019_PRD,Bora_2021_MNRAS,Rather_2023_EPJC}.


In most cases, the stellar stability against radial oscillations
is investigated by applying the Bardeen-Thorne-Meltzer (BTM)
criterion \citep{1966ApJ...145..505B}. The BTM criterion is
usually expressed as follows: when moving toward the direction of
increasing central pressure along the mass-radius curve, at each
extremum, one previously stable radial mode becomes unstable if
the curve bends counterclockwise, while one previously unstable
radial mode becomes stable if the curve bends clockwise. For bare
strange quark objects which include the whole bare strange
planet-bare strange star series, the conclusion on the stability
is quite clear. All configurations before the mass-radius curve
reaches its maximum are stable, with stars of higher central
pressure unstable. It means that all the bare strange planets and
bare strange dwarfs are stable.

However, in the cases of crusted strange quark objects, things become more complicated.
At first glance, strange dwarfs seem to be unstable according to the BTM criterion (see
the right panel of Figure~\ref{Deb_2017_AnPhy}). By contrast, \cite{Glendenning_1995_ApJ}
solved the Sturm-Liouville problem governing stellar stability and claimed that strange
dwarfs are in fact stable. They found that all the eigenvalues of the radial oscillation
mode are positive. It is argued that the strange quark core stabilizes the strange dwarf.
Nevertheless, the detailed mechanism on how the strange quark core stabilizes the
strange dwarf is not addressed. Later, \cite{Alford_2017_ApJ} revisited the problem
and solved the Sturm--Liouville problem again. They found that the lowest eigenvalue
of the radial oscillation mode is negative, which means strange dwarfs are unstable.
They argued that the lowest eigenvalue was essentially omitted by \cite{Glendenning_1995_ApJ}.
Recently, \cite{DiClemente_2023_A&A} and \cite{Goncalves_2023_EPJA} further examined the
issue and found that the difference between \cite{Glendenning_1995_ApJ}
and \cite{Alford_2017_ApJ} is due to the different matching condition used at the interface
between strange quark core and nuclear crust. \cite{Alford_2017_ApJ} have used the so-called
rapid conversion condition, while the calculations of \cite{Glendenning_1995_ApJ} correspond
to the slow conversion condition. As a result, \cite{DiClemente_2023_A&A}
and \cite{Goncalves_2023_EPJA} concluded that strange dwarfs are ``slow-stable'' --- being
stable only when the phase transition process between strange quark matter and nuclear matter is
slower than the radial perturbation.

In fact, the term ``slow-stable'' is generally used to describe hybrid stars, where
the central quark matter is encompassed by nuclear matter and the two kinds of matter
can transfer to each other through phase transition. However, in the context of
crusted strange dwarfs and planets, the crust and strange core are separated by a
strong electric field. There is a ``gap'' (of several hundreds fermis) between strange
quark matter and nuclear matter so that the two ``phases'' do not contact with each
other. As a result, inside strange dwarfs and strange planets, phase transition essentially
cannot proceed between the quark matter and the hadronic matter at the bottom of the crust.
In other words, these light strange quark objects generally satisfy the slow conversion
condition and they are in the ``slow-stable'' state. To conclude, the electric field
between strange core and the crust guarantees the stability of strange dwarfs and strange
planets against usual radial perturbations. However, note that a too large perturbation may
still be able to cause the crust of a strange dwarf to collapse.

\subsubsection{non-radial oscillations}

Non-radial oscillations of neutron stars was initially
studied by \cite{Thorne_1967_ApJ}, which are especially important
for gravitational wave emissions \citep{Price_1969_ApJ}. For
simplicity, we adopt the Cowling approximation
\citep{Cowling_1941_MNRAS,McDermott_1988_ApJ} and ignore the
gravitational perturbations in spacetime. Denoting the deviation
of the fluid element from its equilibrium position as $\xi^i$, we
have \citep{Sotani_PRD_2011},
\begin{eqnarray}
    \xi^i = (e^{-\eta /2} W, -V\partial_\theta, -V sin^{-2} \theta \partial_\phi) r^{-2} Y_{lm} ,
\end{eqnarray}
where $i=r, \theta, \phi$ in this subsection, $W$ and $V$ are
perturbation functions of $t$ and $r$, and $Y_{lm}$ is the
spherical harmonics. The perturbation of the four-velocity $\delta
u^a$ can then be written as
\begin{eqnarray}
    \delta u^a = (0, e^{-\eta /2} \partial_t W, -\partial_t V\partial_\theta, -\partial_t V sin^{-2} \theta \partial_\phi) r^{-2} e^{-\phi} Y_{lm} .
\end{eqnarray}
Note that the energy momentum tensor satisfies
\citep{Sotani_PRD_2011,Curi_2022_EPJC}
\begin{eqnarray}
    \delta (\nabla_b T^{ab} ) = \nabla_b \delta T^{ab} = 0 .
\end{eqnarray}

Assuming that the perturbations are harmonic functions of
time, i.e. $W(r,t) = W(r)e^{i\omega t}$ and $V(r,t)
=V(r)e^{i\omega t}$, the oscillation equations in the Cowling
approximation can be simplified as
\begin{align}
    W' & = \frac{d\epsilon }{dp}[\omega^2 r^2 e^{\eta /2 -\sigma} V+ \frac{1}{2}\sigma' W] - l(l+1)e^{\eta /2}V , \nonumber \\
    V' & = \sigma' V - e^{\eta /2} \frac{W}{r^2}  .
\end{align}
To solve these equations, we again need to specify two boundary
conditions. First, the Lagrangian perturbation of pressure should
vanish at the stellar surface, i.e.
\begin{eqnarray}
    \Delta p = \omega^2 r^2 e^{\eta /2 -\sigma} V+ \frac{1}{2}\sigma' W = 0 .
\end{eqnarray}
Second, the perturbation functions of $W(r,t)$ and $V(r,t)$
satisfy
\begin{align}
    W(r) & = C r^{l+1} + \mathscr{O}(r^{l+3}) , \nonumber \\
    V(r) & = C r^l/l + \mathscr{O}(r^{l+2}) ,
\end{align}
at the star center ($r=0$), where $C$ is a constant.

Solving this eigenvalue problem, we can determine the
characteristic frequency $\omega$ and the corresponding
characteristic function $\xi^i$ associated with the non-radial
oscillations. The solutions are classified into distinct
oscillation modes according to the nature of the characteristic
frequencies and eigenfunctions, which mainly include: 

f-modes (fundamental modes): This is the lowest order mode
of non-radial oscillations. Their frequencies are typically high.
They are primarily driven by the global deformation of fluid
dynamics, reflecting the overall deformation of the star. 

p-modes (pressure modes): For these modes, the frequency
is typically high and it increases with the increasing mode order.
They are primarily driven by pressure waves (sound waves) in the
fluid, reflecting the pressure distribution and the speed of sound
inside the star. 

g-modes (gravity modes): The frequency is usually low.
They are primarily driven by buoyancy forces, reflecting the
density gradients and thermal gradients inside the star. 

Note that under the Cowling approximation, the non-radial
oscillation equations cannot effectively describe the
$\omega$-modes (gravitational wave modes) due to the ignoring of
the metric perturbations. $\omega$-modes are driven solely by
gravitational waves, reflecting the spacetime undulations instead
of material movements. To investigate the gravitational emissions
accurately, metric perturbations in the framework of General
Relativity should be considered and the coupling between the fluid
dynamics equations and the Einstein's field equations should be
solved
\citep{Kokkotas_1992_MNRAS,Andersson_1996_PRL,Andersson_1998_MNRAS}.

\section{hybrid stars}
\label{HybridStars}

A compact star is conceptually divided into five parts,
the atmosphere, the outer crust, the inner crust, the outer core,
and the inner core \citep{Weber_2005_PrPNP}. The atmosphere is a
thin layer of plasma, typically several centimeters in thickness.
The outer crust is composed of atomic nuclei and free electrons,
with the density being lower than $10^{11}$ g/cm$^3$. In the inner
crust, the density becomes higher but is still less than $10^{14}$
g/cm$^3$. Atomic nucleus are disintegrated and neutrons overflow
from the nucleus. In the outer core, the density increases to
exceed the nuclear saturation density so that the matter is
composed of neutrons, protons, electrons, and muons. In this
region, all plasmas are strongly degenerate, with electrons and
muons behaving like ideal Fermi gases, while protons and neutrons,
among other fermionic fluids, may exist in a state of
superfluidity or superconductivity. Many-body nucleon interaction
models are necessary to solve for the equation of state of matter
in this section under the conditions of beta equilibrium and
electric neutrality. When the compact star is massive enough, it
will have a special region at the center, the inner core. In this
case, the density and pressure will be high enough to transform
hadronic matter into quark matter (either with or without strange
quarks). The transition may occur at several times of the nuclear
saturation density and quark matter could coexist with hadronic
matter at around the transition region. We call these stars hybrid
stars. Whether a strange star or a hybrid star is more stable can
be judged by comparing the energy per baryon, which is defined as
the ratio of energy density $\epsilon$ to baryon number density
$n_B$
\begin{eqnarray}
    E/A = \frac{\epsilon}{n_B} .
\end{eqnarray}

Currently, there is no ideal theory that can satisfactory
describe the hadronic phase and the quark phase jointly. So,
different models are employed to describe hadronic phase and quark
phase separately. The two phases are then connected at the
transition region, trying to match with each other through a
particular construction, i.e. either the Maxwell construction or
the Gibbs construction. Several popular models widely used to
describe quark matter have been introduced in Section \ref{InternalStructure}.
Similarly, we have many models for hadronic matter, including the
relativistic mean field (RMF) models
\citep{Walecka_1975_plb,Alaverdyan_2009_ap,Dutra_2014_prc} and the
Brueckner-Hartree-Fock (BHF) approaches
\citep{Li_2010_PhRvC,Li_2012_PhRvC,Tong_2022_apj}. For example,
NL3, TM1, DD-ME2 and FSU Gold models are popular RMF models, while
the Akmal-Pandharipande-Ravenhall (APR) model
\citep{Akmal_1998_prc,Gusakov_2005_MNRAS,Schneider_2019_prc}, is a
typical BHF approach. 

Here we will focus on the transition from the hadronic
phase to the quark phase. Due to unknown physics, the transition
could be a smooth crossover, a sharp first-order transition
involving a latent heat, or even a critical point signaling a
second-order phase transition. Two methods have been engaged to
math the two phases at the transition region, i.e. the Maxwell
construction or the Gibbs construction. 

\subsection{Maxwell construction}

\begin{figure}[htbp]
    \includegraphics[width=\textwidth]{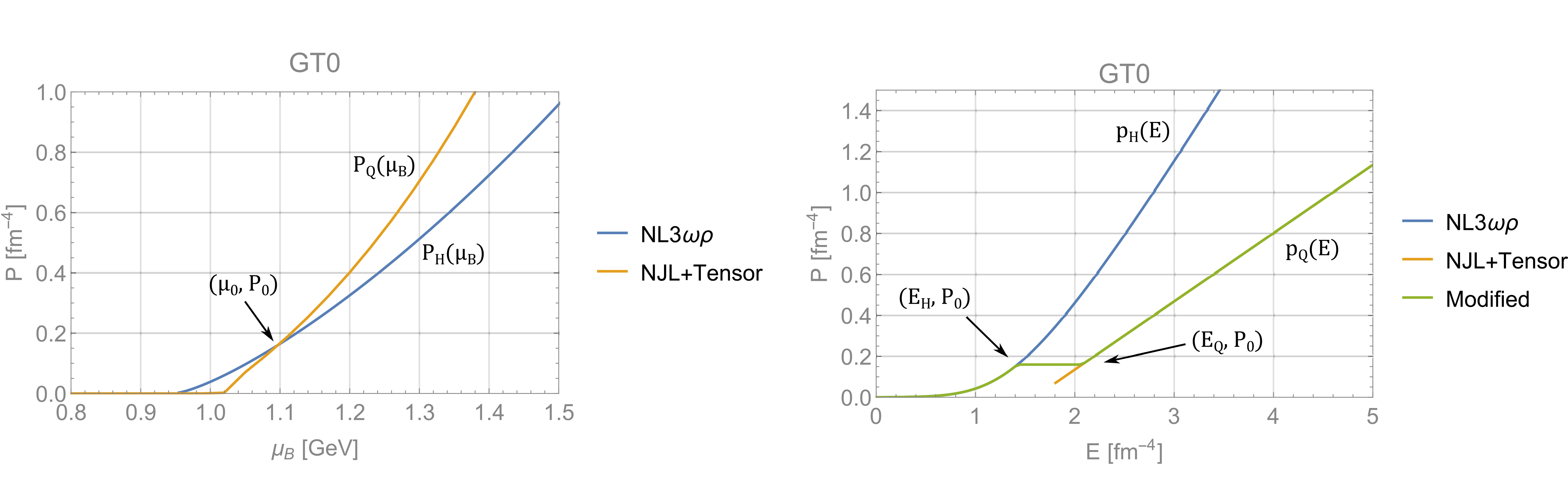}
    \centering
    \caption{The Maxwell construction between hadronic matter and quark
    matter \citep{Matsuoka_2018_prd}. The hadronic phase (NL3$\omega \rho$ model)
    and quark phase (NJL model) are represented by blue lines and orange lines,
    respectively. The green line illustrates the overall EOS. GT0 in the figure
    means the tensor interaction is zero.}
    \label{Matsuoka_2018_prd}
\end{figure}

The Maxwell construction describes a fist-order transition
from hadronic matter to quark matter. In this case, the baryon
number is conserved and the two phases are in equilibrium. There
could be a sharp interface between the two phases so that the
transition is a definite phase transitio, as shown in the left
panel of Figure \ref{Matsuoka_2018_prd}. When the pressure ($P$)
is higher than the critical point ($P_0$), the matter is in the
quark configuration, while when $P$ is lower than $P_0$, the
matter is in the hadronic phase. To avoid the instabilities caused
by the long-range nature of electromagnetic forces, each of the
two phases have to maintain electric neutrality independently,
i.e. 
\begin{eqnarray}
    q^{quark} = q^{hadron} = 0 .
\end{eqnarray}

Under this fist-order transition, three equilibrium
conditions should also be satisfied, i.e. the chemical potential
equilibrium 
\begin{eqnarray}
    \mu_B^{quark} = \mu_B^{hadron} = \mu = \mu_n ,
\end{eqnarray}
the mechanical equilibrium
\begin{eqnarray}
    P^{quark} (\mu, T) = P^{hadron} (\mu, T) ,
\end{eqnarray}
and the thermal equilibrium
\begin{eqnarray}
    T^{quark} = T^{hadron} = T ,
\end{eqnarray}
where $\mu_B$ is the chemical potential of baryons, $\mu$
is the chemical potential of the whole system, $\mu_n$ is the
chemical potential of neutron, and $T$ is temperature. The
chemical potential equilibrium and mechanical equilibrium
conditions are also reflected in Figure \ref{Matsuoka_2018_prd}. 

The overall EOS of a hybrid star is shown in the right
panel of Figure \ref{Matsuoka_2018_prd}. There is a ``plateau'' in
the EOS curve, which corresponds to the transition between the
hadronic phase and the quark phase. Note that the pressure is a
continuous function inside the star, but there is a discontinuity
in the energy density at the transition point. Such a jump in the
energy density is also known as ``latent heat'', which is a
hallmark characteristic of a first-order phase transition. 

\subsection{Gibbs construction}

The Gibbs construction is widely employed to describe the
complex mixed matter inside compact stars
\citep{Glendenning_1992_PhRvD}. For the ``complex'' mixed phase
possibly existed inside hybrid stars, keeping local electric
neutrality independently is unreasonable, because the particles
can interact with each other in a complicated way. Therefore,
charge is conserved and electric neutrality is maintained only for
the whole system, but not for each phase. This is a much weaker
constraint comparing to that in the Maxwell construction. 

According to the Gibbs construction, the chemical
potential of each component is still equal between different
phases, i.e.
\begin{eqnarray}
    \mu_B^{quark} = \mu_B^{hadron} = \mu = \mu_n ,\\
    \mu_Q^{quark} = \mu_Q^{hadron} = \mu_e ,
\end{eqnarray}
where the subscript $B$ and $Q$ represent the baryon phase and
quark phase, respectively. $\mu_e$ is the chemical potential of
electrons, which appears only in the quark phase. For the pressure
and temperature , we also have
\begin{eqnarray}
    P^{quark} (\mu, T) = P^{hadron} (\mu, T) ,
\end{eqnarray}
\begin{eqnarray}
    T^{quark} = T^{hadron} = T.
\end{eqnarray}
For simplicity, we could take the temperature as zero, i.e. $T =
0$. In this way, hadrons and quarks can not only coexist but also
mix together in hybrid stars. So, the Gibbs construction is more
flexible than the Maxwell construction.

The global electronic neutral condition is expressed as
\begin{eqnarray}
    \chi q^{quark} + (1-\chi) q^{hadron} = 0 ,
\end{eqnarray}
where $\chi$ is the volume fraction of quarks that satisfies $0
\le \chi \le 1$. In the mixed region, we have
\begin{eqnarray}
    \chi = \frac{V^{quark}}{V^{quark}+V^{hadron}} .
\end{eqnarray}
Then the energy density of the mixed region is
\begin{eqnarray}
    \epsilon^{mixed} = \chi \epsilon^{quark} + (1-\chi) \epsilon^{hadron} ,
\end{eqnarray}
and the baryon number density is
\begin{eqnarray}
    \rho^{mixed} = \chi \rho^{quark} + (1-\chi) \rho^{hadron} .
\end{eqnarray} 

\begin{figure}[htbp]
    \includegraphics[width=0.5\textwidth]{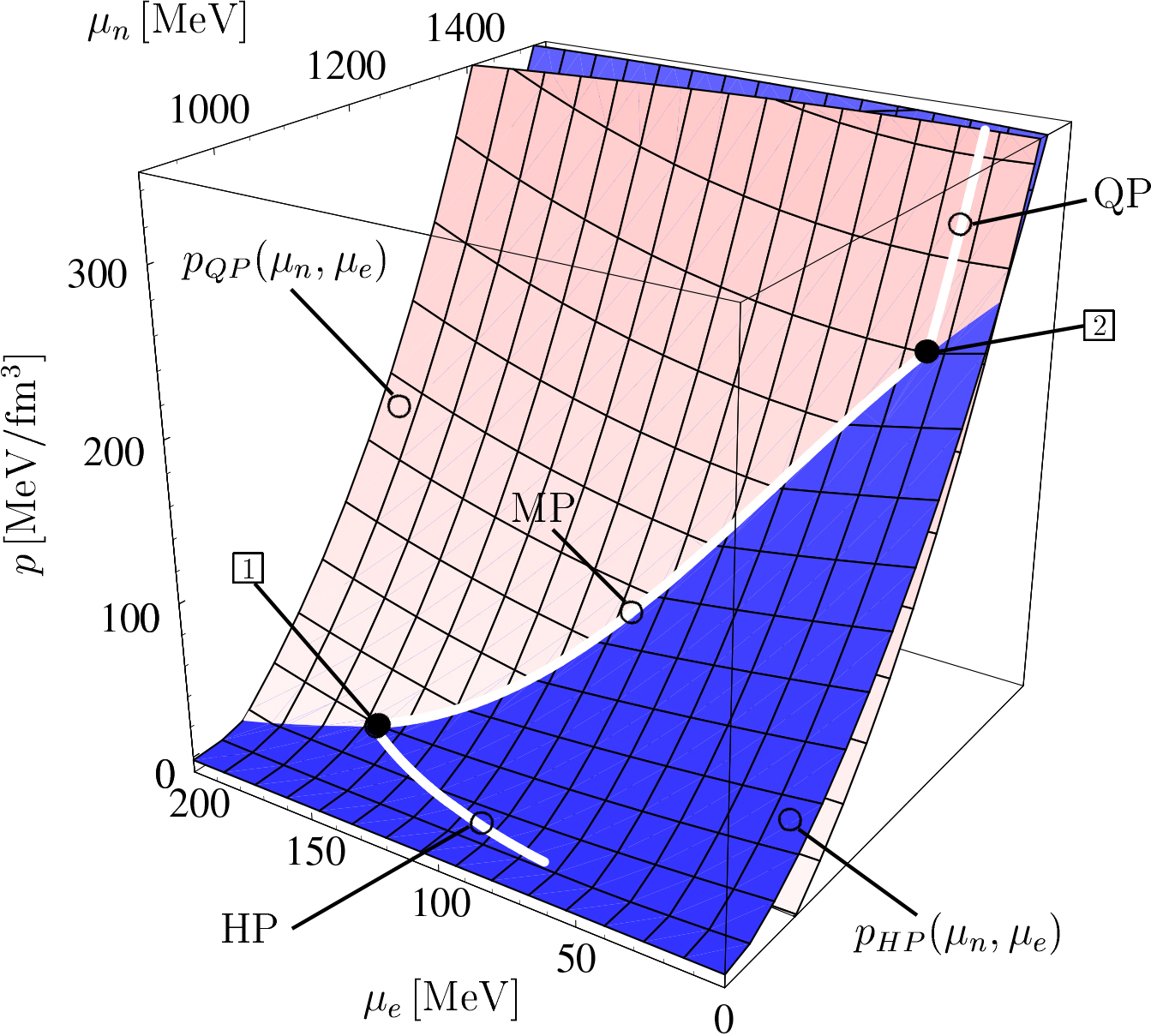}
    \centering
    \caption{The Gibbs construction between hadronic matter and quark
    matter \citep{Schertler_2000_NuPhA}. The pressure of the hadronic
    phase ($p_{HP}$) and the quark phase ($p_{QP}$) is plot as a
    function of the two independent chemical potentials of $\mu_n$
    and $\mu_e$. The intersection curve of the two pressure planes
    corresponds to the Gibbs condition, where the mixed phase exist.
    The white lines of $HP$ and $QP$ on the pressure surfaces show the
    pressure of the hadronic phase and the quark phase under the
    condition of charge neutrality, respectively.
     }
    \label{Schertler_2000_NuPhA}
\end{figure}

Comparing with the fixed parameters in the Maxwell
construction, the parameters in the Gibbs construction are decided
by the percentage of hadrons and quarks. In this case, the energy
density is a continuous function and the conversion is smooth
rather than a sharp phase boundary existed in the Maxwell
construction. It is possible that there is no critical point of
any phase transition, but only a crossover from hadronic matter to
quark matter in the transition region. Figure
\ref{Schertler_2000_NuPhA} illustrates a typical Gibbs
construction between hadronic matter and quark matter. The
hadronic EOS used in the figure is from \cite{Ghosh_1995_ZPhyA}
and the quark EOS is the effective mass bag model
\citep{Schertler_1997_NuPhA,Schertler_1998_NuPhA}. The pressure is
plot as the function of two independent chemical potentials
($\mu_n$, $\mu_e$). The intersection curve of the two pressure
surfaces shows the solution of the Gibbs condition, where the
mixed phase presents. In this transition region, all physical
parameters change continuously and the free energy of the mixed
phase is at a minimum. 

The Gibbs construction is originally utilized to delineate
the multi-phase equilibrium. The chemical potentials of each
species are equal in the two phases when they coexist, which
satisfies the fundamental requirement for thermodynamic
equilibrium. It is not only pertinent to describe first-order
phase transitions, but also can be employed to depict higher-order
phase transitions and crossover phenomena under suitable
circumstances. 

\section{GW emission from strange stars}
\label{GW}

GW emission was first proposed by Einstein as a prediction of the General Theory of
Relativity \citep{Einstein_1916_SPAW,Einstein_1918_SPAW}. Any changes in the
distribution of matter may lead to the variation of the curvature of space-time,
causing energy to be carried away in the form of gravitational waves.
The detection of GW signals
in 2015 by the LIGO collaboration \citep{Abbott_2016_PhRvL} marks the beginning of
a new multi-messenger era in astronomy. Many efforts have been made to explore various
possible mechanisms that could generate GWs efficiently.  Compact stars, due to their
extreme density and dynamic motion, serve as crucial sources of GWs. In this aspect,
GW emission associated with strange stars may have some special features since their
internal composition and structure are different from normal neutron stars. We thus
could potentially use GW observations to help identify strange stars.

\subsection{GWs from binary strange star systems}

Coalescence of binary systems is the most significant stellar GW sources. BH-BH,
BH-NS, and NS-NS binary systems are common GW sources. Here we focus on
binaries containing strange quark stars, such as BH-SQS and SQS-SQS systems. 
The detection of the controversial GW190814 event, which involves a possible mass-gap
compact object ($2.5$ -- $2.67 M_\odot$), has drawn a wide attention in the community.
Although most researchers believe it could be a neutron star, the possibility that it is a
strange star has also been discussed in numerous
studies \citep{I.Bombaci_2021_PRL,Miao_2021_ApJL,ZachariasRoupas_2021_PRD,Lopes_2022_ApJ,Oikonomou_2023_PhysRevD}.
Interestingly, two frequently mentioned GW events, GW170817 and GW190425, are
suggested to originate from SQS-SQS systems by \cite{Miao_2021_ApJL},
\cite{AnilKumar_2022_MN}, and \cite{Sagun_2023_ApJ}. Furthermore, theoretical
calculations of GW radiation from SQS-SQS binary systems have been preformed
by \cite{Limousin_2005_PhRvD}, \cite{Gondek-Rosinska_2008_arXiv},
and \cite{Bauswein_2010_PhRvD}. \cite{Moraes_2014_MNRAS} even
studied the possible existence of NS-SQS binaries and investigated their GW emission.

\begin{figure}[htbp]
    \includegraphics[scale=.45]{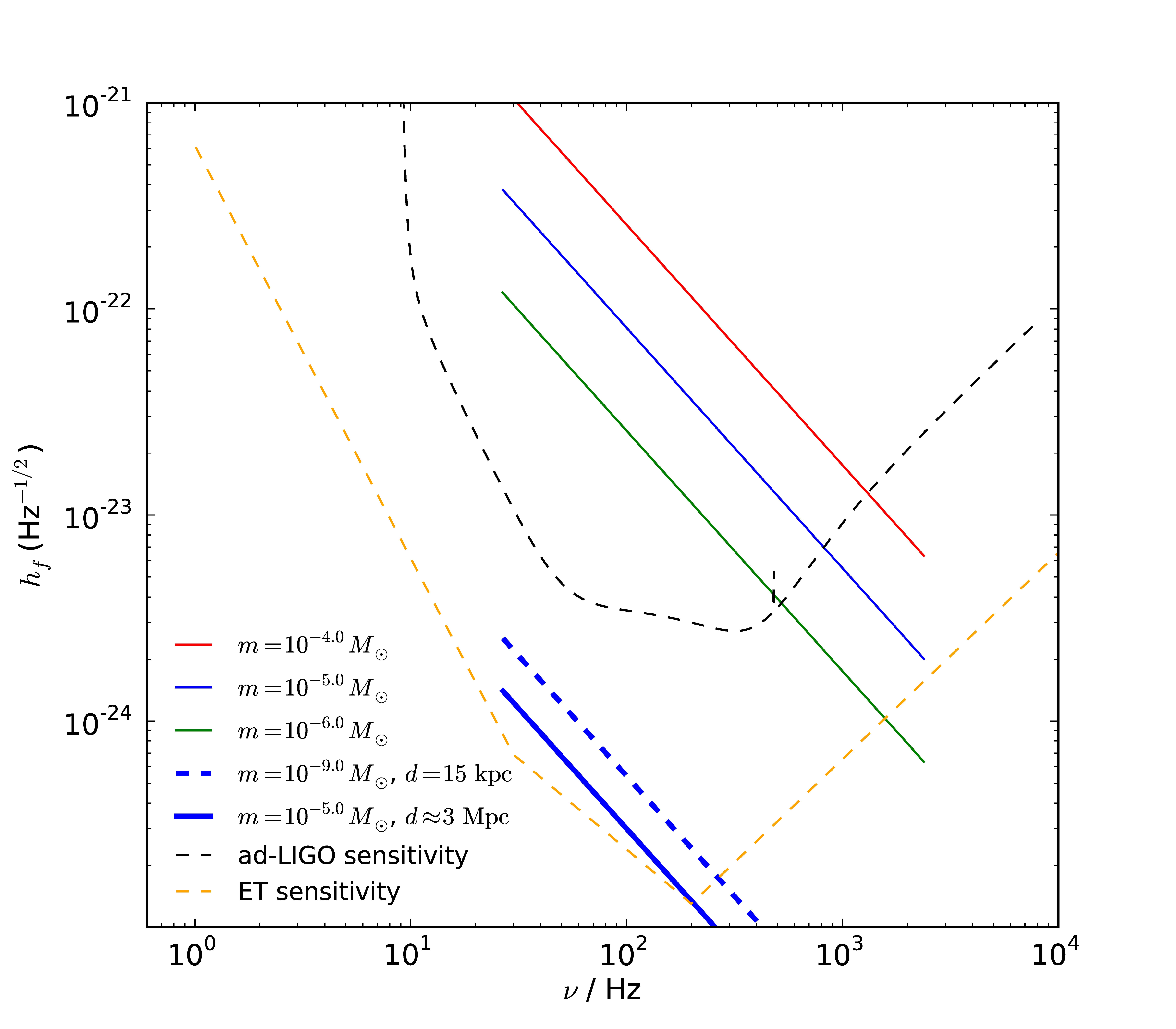}
    \centering
    \caption{Strain spectral amplitude of GWs against frequency for coalescing
                     SQS-strange quark planet systems  \citep{Geng_2015_APJL}. Various masses are assumed for the
                     strange quark planet, and different distances are taken for the system. The
                     sensitivity curves of advance-LIGO and ET are also shown for comparison.  }
    \label{Geng_2015_APJL}
\end{figure}

\begin{figure}[htbp]
    \includegraphics[scale=.35]{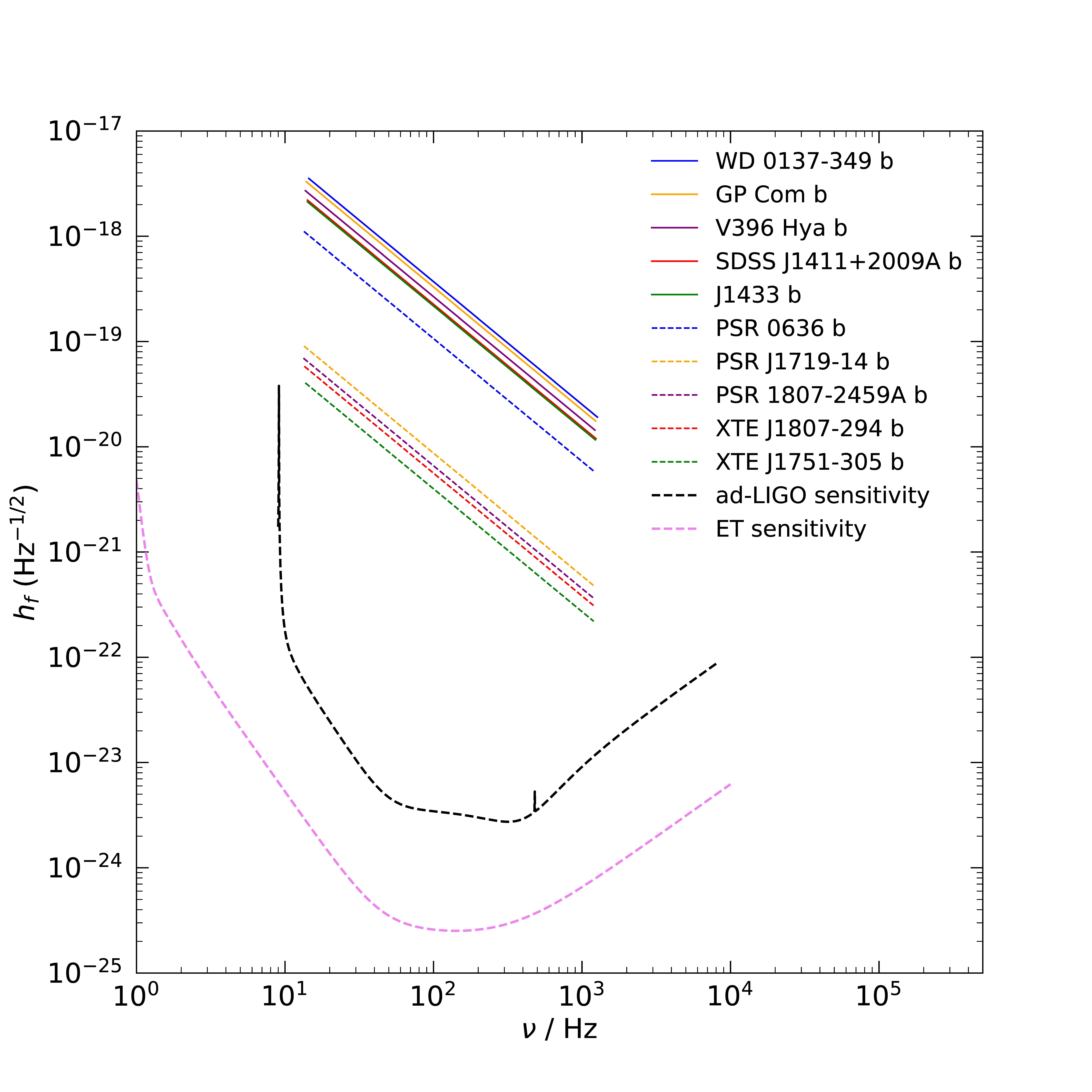}
    \centering
    \caption{Strain spectral amplitude of GWs against frequency for coalescing strange
                    quark matter planetary systems \citep{Kuerban_2020_ApJ}. The names of the candidate
                    strange quark planets are marked in the plot. This figure shows the strain spectral
                    amplitude of the GW emission when they finally merge with their host in the future. }
    \label{Kuerban_2020_ApJ}
\end{figure}

It should be noted that binary systems containing low-mass strange quark objects can
also be strong GW sources. For instance, \cite{Lv_2009_RAA} investigated GW emission
from a white dwarf (WD)--strange dwarf system. Moreover, \cite{Perot_2023_PhRvD}
argued that GW signal could serve as a better probe to distinguish between strange
dwarfs from white dwarfs in binaries. More interestingly, strange quark planets could also
stably exist. It is proposed that the merger of a strange quark planet with a SQS
can also lead to strong GW emission \citep{Geng_2015_APJL,Zhang_2024_MNRAS}.
Generally, GW emission from merging SQS-strange quark planet will be too weak
to be detected if it happens at cosmological distances. However, such events occurring
in our Galaxy or in nearby local galaxies is detectable for the Advanced
LIGO \citep{Harry_2010} and Einstein Telescope \citep{Hild_2008arXiv}.
Figure \ref{Geng_2015_APJL} shows that the strain spectral amplitude of GWs from
coalescing SQS–strange quark planet systems is well above the detection limit
when they happen in our Galaxy or in local galaxies. At the same time, we could
also try to identify strange quark objects by searching for close-in planets around
pulsars. The period of normal matter planet around a pulsar cannot be less than
6100 s, since it would be tidally disrupted in such a close orbit. On the contrary,
the period of strange quark planet could be much less than 6100s due to its extreme
high density. Using this method, several extrasolar planetary systems that contain
a close-in planet have been argued to be possible candidates of strange planetary
systems \citep{Kuerban_2020_ApJ}. Since these extrasolar planetary systems
are all relatively close to us, Figure \ref{Kuerban_2020_ApJ} shows that if they merge,
the GW emission will be well above the detection limit of current and future GW
experiments \citep{Kuerban_2020_ApJ}.

\begin{figure}[htbp]
    \includegraphics[width=.55\textwidth]{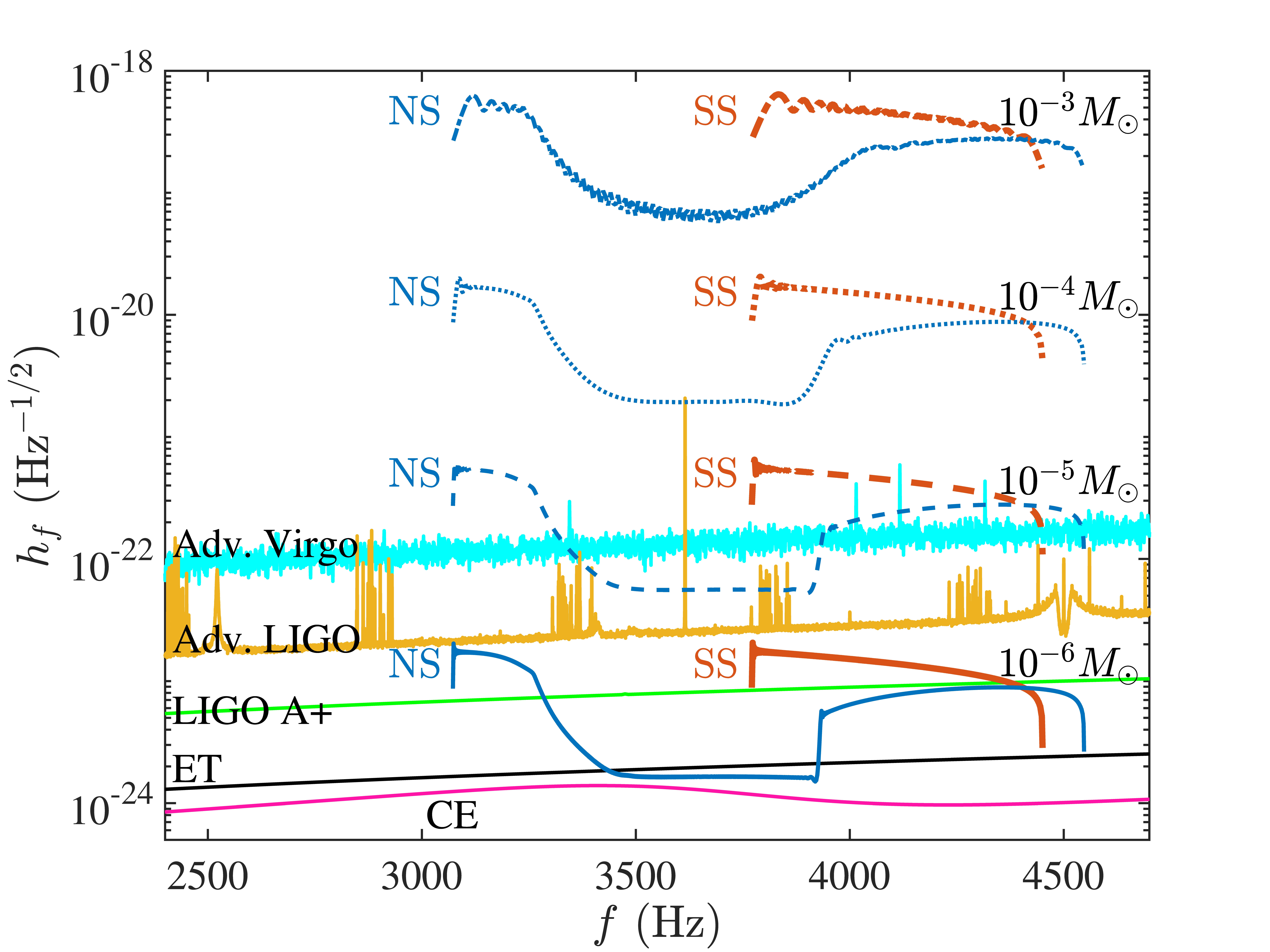}
    \centering
    \caption{Strain spectral amplitude of GWs against frequency for a primordial black
                   hole inspiraling inside a strange star or a neutron star \citep{Zou_2022_ApJ}.
                   The system is assumed to be at 1 kpc from us. The sensitivity curves of several
                   GW experiments are shown for comparison. }
    \label{Zou_2022_ApJ}
\end{figure}

Recently, \cite{Zou_2022_ApJ} studied the GW emission produced when a primordial black hole
inspirals inside a strange star.  The black hole will grow when it swallows the matter from the
strange star. It will finally fall toward the center of the strange star and convert the whole star into
a stellar mass black hole. During the process, strong GWs will be emitted, whose frequence falls
in the range of various ground-based GW detectors, such as the Advanced Virgo, Advanced LIGO,
LIGO A+ upgrade, Einstein Telescope (ET), and Cosmic Explorer (CE). More importantly, the
GW signals will be different from that produced when a primordial black hole inspirals inside a
neutron star, as illustrated in Figure \ref{Zou_2022_ApJ}. Observation of such GW events thus
could provide a useful discrimination between strange stars and neutron stars.

\subsection{GWs from other mechanisms concerning strange stars}

Aside from merger events, the collapse of a neutron star induced by a phase transition to a strange star can also
lead to strong GW emission. An increase in the central density of a neutron star can trigger a
phase transition from hadronic matter to deconfined quark matter within the core. This transition
may lead to the collapse of the whole neutron star into a more compact strange star, accompanied
by the emission of GWs. \cite{Lin_2006_ApJ} and \cite{Abdikamalov_2009_MNRAS} utilized
hydrodynamic simulations to investigate the phase transition process and computed the GW emissions.
Recently, \cite{Yip_2023_arXiv} incorporated magnetic fields into their numerical studies to explore
the formation of a magnetized strange star and computed the GW signals through general relativistic
magnetohydrodynamics simulations. These studies reveal that the emitted GW spectrum is primarily
dominated by two fundamental modes: the quasi-radial F mode ($l = 0$) and the quadrupolar $^2 f$
mode ($l = 2$). Additionally, \cite{Yip_2023_arXiv} demonstrated that observations of these
fundamental modes can help measure the magnetic field strength of the interior toroidal field
and the baryonic mass fraction of matter in the mixed phase.

In addition to transient GWs produced from catastrophic events, continuous GW emission
can also arise from global oscillations of strange stars. Specifically, r-mode oscillations occurring
in rotating SQSs \citep{Andersson_2002_MNRAS,Rupak_2013_PhRvC,Wang_2019_RAA} are
potential mechanisms to produce continuous GWs. The r-mode instability leads to a gradual
loss of angular momentum from the compact star, causing it to spin down and emit continuous
GWs. Additionally,  \cite{Gondek_2003_A&A} proposed that a triaxial, ``bar shaped'' strange star
could be an efficient source of continuous GW radiation, which is called the bar-mode GW
emission. Continuous GW emissions are generally much weaker as compared with that of the catastrophic
GW events \citep{2022Univ....8..442Z}. However, with the improvement in the sensitivity
of future detectors, GWs from r-mode and bar-mode instabilities may be detectable, which
will provide a novel tool for probing the dense matter in compact stars.

The number of detected GW events is increasing rapidly in recent years. Although most of the observed
GW events are produced by mergering binary black holes and are not directly related to neutron
stars/strange stars, it is possible that more and more GW events involving neutron stars/strange stars
will be detected in the near future. GW observations will be a powerful tool to reveal the internal
composition and structure of pulsars.

\section{Electromagnetic bursts from strange stars}
\label{ElectromagneticBursts}

With a strong gravity and magnetic field, a strange star can accrete matter from the surrounding medium
or from a companion star. This will lead to some kinds of electromagnetic bursts, such as gamma-ray bursts (GRBs)
and fast radio bursts (FRBs).

\subsection{GRBs from strange stars}
\label{GRB}

GRBs are one of the most violent stellar explosions. The isotropic equivalent $\gamma$-ray energy released in a typical
GRB is in a range of $10^{50}$ -- $10^{53}$ ergs. GRBs can be classified into two categories according to
their duration, i.e. long GRBs that last for longer than $\sim$ 2 s and short GRBs shorter than $\sim$ 2 s. After more than 50
years of study, it is now generally believed that long GRBs are produced by the collapse of massive stars, while short
GRBs are produced by the merger of binary compact stars. However, the possibility that some GRBs are produced
by other mechanisms still cannot be expelled \citep{Levan_2016_SSR,Zou_2021_ApJ}.
For example, strange stars could be involved in some of these fierce bursts.

The conversion of neutron stars to strange stars through a phase transition process will lead to the release of
a huge amount of energy, which would be large enough to produce short
GRBs \citep{Cheng_1996_PhRvL,Wang_2000_A&A,Bombaci_2000_ApJ,Shu_2017_MPLA,Prasad_2018_ApJ}.
Many factors can trigger the phase transition process. First, when a neutron star accretes matter from the
ambient environment, its mass increases and will finally exceed a maximum value, beyond which
the whole star will collapse and be transferred to a strange star. For example, \cite{Berezhiani_2003_ApJ}
argued that the central object produced in supernova explosion may be metastable.  It could
accrete the fall-back matter and collapse to form a strange star.  Second, nuclear reactions inside neutron
stars can change their internal structure. When the pressure, density and temperature are high enough,
neutron matter can transfer to form quark matter. \cite{Drago_2004_EPJA} considered color
superconductivity in strange stars and found that diquark condensate could occur, which further
increases the energy release during the conversion process. Thirdly, sudden fluctuations of density
inside neutron stars can create a ``seed'' of strange quark matter, which triggers the phase transition
and propagates outward to the stellar surface. \cite{Mallick_2014_NuPhA} found that during the
spin down process, the central density of a massive neutron star may increase significantly, triggering
a phase transition. The effect of magnetic field in the process is also considered.   When a neutron star
is converted into a strange star, the energy released during the process is of the magnitude
of $\sim 10^{53}$ ergs, sufficient enough to power cosmological short GRBs.

A strange star can be covered by a normal matter crust. The collapse of the crust can also release
a large amount of energy and produce an electromagnetic burst. According to
Equation \ref{funcMIT}, the density is a finite value for strange quark matter when the pressure is
zero, indicating that it is self-bound. As a result, the mass of strange stars can have a very wide
range, i.e. from planetary mass strange objects to nearly two-solar-mass strange stars.
On the surface of strange quark matter, quarks are confined by short-range strong interactions, while
electrons are confined by long-range electromagnetic interactions. It leads to the formation of
an electric field on a lengthscale of hundreds of fermis. The intensity of the electric field can be
up to $10^{17}$ V cm$^{-1}$. Due to the presence of this strong electric field, normal nuclear
matter are expelled and accumulates over the surface to form a crust \citep{Alcock_1986_ApJ,Huang_1997_A&A}.
Numerical simulations by \cite{Jia_2004_ChA&A} shows that when a strange star accretes matter,
the crust will collapse to trigger a short burst. If the accretion continues, the crust will be re-built
and re-collapse again and again, resulting in periodic explosive activities. This mechanism may
account for some kinds of soft gamma-ray repeaters.

\subsection{FRBs from strange stars}

FRBs are fast radio bursts that happen randomly from the sky, typically lasting
for a timescale of milliseconds \citep{Lorimer_2007_Science,Thornton_2013_Science}.
Despite their short durations, the energy released during the burst is immense.
About seven hundred FRB sources have been discovered to date, with nearly 30
of them confirmed to exhibit repeating explosive activities. However, the possibility
that other FRB sources may also be repeating still cannot be expelled yet.
Comprehensive statistical analyses on the properties of repeating FRBs have been
extensively carried out based on current astronomical observations, imposing various
constraints on their nature \citep{Li_2017_RAA,Li_2021_ApJ,Hu_2023_ApJS},  but the
origin of FRBs still remains an open problem in need of further investigations.

Various models have been proposed to explain the observational characteristics of FRBs.
For non-repeating FRBs, \cite{Geng_2015_ApJ} and \cite{Geng_2020_ApJ} suggested that they could
result from the collision between a compact star and an asteroid.  Subsequently, for repeating
bursts, researchers went further to argue that they may arise from multiple collisions as
magnetized NSs travel through asteroid belts \citep{Dai_2016_ApJ}. On the other hand,
\cite{Kurban_2022_ApJ} and \cite{Nurmamat_2024_EPJC} proposed that repeating bursts may originate
from the tidal interactions in a highly elliptical planetary system, in which either a
neutron star or a strange star could be involved. In their framework, a planet moves in a
highly elliptical orbit around the compact star. The planet will be partially disrupted every
time it passes through the periastron since it is very close to the compact host star,
generating smaller clumps that finally collide with the host to produce periodically
repeating FRBs. These models can satisfactorily explain many of the observed features
of repeating FRB sources.

\begin{figure}[htbp]
    \includegraphics[scale=1.0]{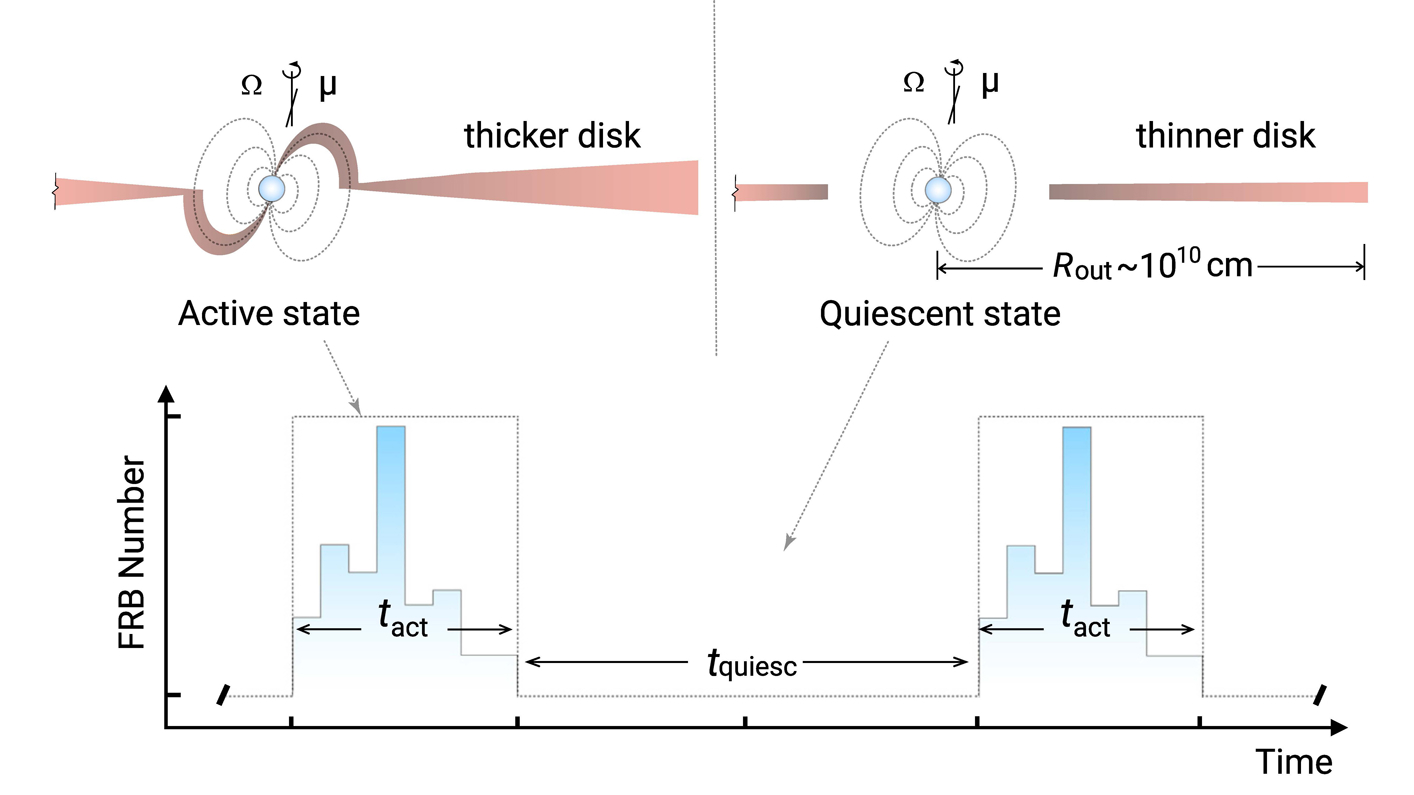}
    \centering
    \caption{Schematic illustration of periodic repeating fast radio bursts produced in the strange star crust
                  collapse scenario \citep{Geng_2021_Innov}.  In the active state, the accretion rate is high, leading
                  to frequent burst activities. In the quiescent state, the accretion rate is low, and no burst will be
                  generated. }
    \label{Geng_2021_Innov}
\end{figure}

In the previous subsection, we have mentioned that GRBs could originate from the
collapse of the crust of a strange star. Similarly, it is also plausible that FRBs may be
triggered by such processes, especially when the strange star is a strongly magnetized
object. Enormous energy is released during the collapse, giving birth to a large amount
of electron/positron pairs. The calculations by \cite{Zhang_2018_ApJ} show that these
electron/positron pairs would be accelerated to relativistic velocities far above the polar
cap region of the strange star, streaming out along the magnetic field lines and ultimately
resulting in a short burst in radio waves. More interestingly, \cite{Geng_2021_Innov}
found that the collapse of the crust of a strange star could happen repeatedly, thus
can also serve as a potential mechanism for periodic repeating FRBs. In their framework,
the strange star accretes matter from its companion. The accretion flow streams along the
magnetic field lines and accumulate in the polar cap region. When the matter at the cap
becomes too heavy, the local crust will collapse, triggering an FRB. The crust can be
re-built when the accretion process continues and may collapse again once it is overloaded.
Periodically repeating FRBs are generated in this way. It should be noted that the active
window and quiescent stage in one period could be governed by thermal-viscous instabilities
(see Figure \ref{Geng_2021_Innov}).

\section{Concluding Remarks}
\label{ConcludingRemarks}

Strange quark matter could be the true ground state of matter at
extreme densities. Such a hypothesis need to be tested through
astronomical investigations. Essentially, we should try to
discriminate between neutron stars and strange stars through
observations. In this article, we present a brief review on some
recent progresses in this field. Various models describing quark
confinement are introduced. The corresponding EOS derived from
these models are presented and compared. By combining these EOSs
with the TOV equations, we can calculate the inner structure of
strange stars, deriving the mass-radius relation for the whole
sequence of strange quark objects. The tidal deformability
and the Love number measured through gravitational wave
observations may help diagnose the EOS. Radial and non-radial
oscillations with different modes can also be used to probe the
internal structure of strange stars. The properties of hybrid
stars, in which quarks and hadrons may coexist, are also
discussed.  Some special kinds of electromagnetic bursts could
also be connected with strange stars and can be used as a probe of
these exotic objects.

GW observation is a hopeful tool to test the existence of strange stars. The coalescence of
binary compact systems which includes at lease one strange object could lead to strong GW
emission. Previously, people mainly concentrate on relatively high mass binaries, such
as BH-SQS and SQS-SQS systems. In the past decade, it is found that when a
low mass strange quark planet merge with its host strange star, strong GW emission will
also be generated and could be detectable to us if the merger event happens in our Galaxy
or in local galaxies \citep{Geng_2015_APJL}. Furthermore, strange quark planets revolving
around their host strange star in a close-in orbit could even be persistent GW
sources  \citep{Kuerban_2020_ApJ,Zhang_2024_MNRAS}, which could be potential goals of
space-based GW experiments in the future.
Apart from binary interactions, collapses induced by phase transitions (e.g., from hadronic
matter to deconfined quark matter) and global oscillations are also likely to generate
GW emission, which could also be tested by the next generation GW detectors.

GRBs and FRBs are fierce events possibly connected to strange stars. The merger of double
strange stars can produce a short GRB, which is very similar to that of binary neutron star
coalescence. Additionally, the conversion of neutron stars to strange stars may also act as
the energy sources for short GRBs. The total energy released during this phase transition process
is estimated to be $\sim 10^{53}$ ergs, which is large enough to power short GRBs occurring at
cosmological distances. Furthermore, the collapse of the crust of strange stars could also act as
a special mechanism to release a large amount of energy, which could explain FRBs (either one-off
or repeating) or some weaker GRBs. The collapse may be triggered by accretion process that
makes the crust overloaded. Finally, the collision of an asteroid with a strange star will also lead
to some interesting phenomena, which still need to be investigated in detail in the future.

\section*{Conflict of Interest Statement}

The authors declare that the research was conducted in the absence of any commercial or
financial relationships that could be construed as a potential conflict of interest.

\section*{Author Contributions}

X-LZ: conceptualization, investigation, validation, and writing-original draft.
Y-FH: conceptualization, funding acquisition, validation, writing-review, and editing.
Z-CZ: conceptualization, validation, writing-review, and editing.

\section*{Funding}
The authors declare that financial support was received for the research, authorship, and/or publication of this article.
This study was supported by the National Natural Science Foundation of China (Grant Nos. 12233002),
by National SKA Program of China No. 2020SKA0120300,
by the National Key R\&D Program of China (2021YFA0718500).

\section*{Data Availability Statement}
No new data were generated or analyzed in support of this research.

\bibliographystyle{Frontiers-Harvard} 
\bibliography{ref}

\begin{thebibliography}{136}
\providecommand{\natexlab}[1]{#1}
\expandafter\ifx\csname urlstyle\endcsname\relax
  \providecommand{\doi}[1]{doi:\discretionary{}{}{}#1}\else
  \providecommand{\doi}{doi:\discretionary{}{}{}\begingroup
  \urlstyle{rm}\Url}\fi
\providecommand{\selectlanguage}[1]{\relax}
\providecommand{\bibAnnoteFile}[1]{%
  \IfFileExists{#1}{\begin{quotation}\noindent\textsc{Key:} #1\\
  \textsc{Annotation:}\ \input{#1}\end{quotation}}{}}
\providecommand{\bibAnnote}[2]{%
  \begin{quotation}\noindent\textsc{Key:} #1\\
  \textsc{Annotation:}\ #2\end{quotation}}

\bibitem[{{Abbott} et~al.(2016){Abbott}, {Abbott}, {Abbott}, {Abernathy},
  {Acernese}, {Ackley} et~al.}]{Abbott_2016_PhRvL}
{Abbott}, B.~P., {Abbott}, R., {Abbott}, T.~D., {Abernathy}, M.~R., {Acernese},
  F., {Ackley}, K., et~al. (2016).
\newblock {Observation of Gravitational Waves from a Binary Black Hole Merger}.
\newblock \emph{Phys. Rev. Lett.} 116, 061102.
\newblock \doi{10.1103/PhysRevLett.116.061102}
\bibAnnoteFile{Abbott_2016_PhRvL}

\bibitem[{{Abdikamalov} et~al.(2009){Abdikamalov}, {Dimmelmeier}, {Rezzolla},
  and {Miller}}]{Abdikamalov_2009_MNRAS}
{Abdikamalov}, E.~B., {Dimmelmeier}, H., {Rezzolla}, L., and {Miller}, J.~C.
  (2009).
\newblock {Relativistic simulations of the phase-transition-induced collapse of
  neutron stars}.
\newblock \emph{Mon. Not. R. Astron. Soc.} 392, 52--76.
\newblock \doi{10.1111/j.1365-2966.2008.14056.x}
\bibAnnoteFile{Abdikamalov_2009_MNRAS}

\bibitem[{Akmal et~al.(1998)Akmal, Pandharipande, and
  Ravenhall}]{Akmal_1998_prc}
Akmal, A., Pandharipande, V.~R., and Ravenhall, D.~G. (1998).
\newblock {The Equation of state of nucleon matter and neutron star structure}.
\newblock \emph{Phys. Rev. C} 58, 1804--1828.
\newblock \doi{10.1103/PhysRevC.58.1804}
\bibAnnoteFile{Akmal_1998_prc}

\bibitem[{Alaverdyan(2009)}]{Alaverdyan_2009_ap}
Alaverdyan, G.~B. (2009).
\newblock {Relativistic mean-field theory equation of state of neutron star
  matter and a Maxwellian phase transition to strange quark matter}.
\newblock \emph{Astrophysics} 52, 132--150.
\newblock \doi{10.1007/s10511-009-9043-y}
\bibAnnoteFile{Alaverdyan_2009_ap}

\bibitem[{{Alcock} et~al.(1986){Alcock}, {Farhi}, and
  {Olinto}}]{Alcock_1986_ApJ}
{Alcock}, C., {Farhi}, E., and {Olinto}, A. (1986).
\newblock {Strange Stars}.
\newblock \emph{Astrophys. J.} 310, 261.
\newblock \doi{10.1086/164679}
\bibAnnoteFile{Alcock_1986_ApJ}

\bibitem[{{Alford} et~al.(2017){Alford}, {Harris}, and
  {Sachdeva}}]{Alford_2017_ApJ}
{Alford}, M.~G., {Harris}, S.~P., and {Sachdeva}, P.~S. (2017).
\newblock {On the Stability of Strange Dwarf Hybrid Stars}.
\newblock \emph{Astrophys. J.} 847, 109.
\newblock \doi{10.3847/1538-4357/aa8509}
\bibAnnoteFile{Alford_2017_ApJ}

\bibitem[{{Andersson} et~al.(2002){Andersson}, {Jones}, and
  {Kokkotas}}]{Andersson_2002_MNRAS}
{Andersson}, N., {Jones}, D.~I., and {Kokkotas}, K.~D. (2002).
\newblock {Strange stars as persistent sources of gravitational waves}.
\newblock \emph{Mon. Not. R. Astron. Soc.} 337, 1224--1232.
\newblock \doi{10.1046/j.1365-8711.2002.05837.x}
\bibAnnoteFile{Andersson_2002_MNRAS}

\bibitem[{{Andersson} and {Kokkotas}(1996)}]{Andersson_1996_PRL}
{Andersson}, N. and {Kokkotas}, K.~D. (1996).
\newblock {Gravitational Waves and Pulsating Stars: What Can We Learn from
  Future Observations?}
\newblock \emph{Phys. Rev. Lett.} 77, 4134--4137.
\newblock \doi{10.1103/PhysRevLett.77.4134}
\bibAnnoteFile{Andersson_1996_PRL}

\bibitem[{{Andersson} and {Kokkotas}(1998)}]{Andersson_1998_MNRAS}
{Andersson}, N. and {Kokkotas}, K.~D. (1998).
\newblock {Towards gravitational wave asteroseismology}.
\newblock \emph{Monthly Notices of the Royal Astronomical Society} 299,
  1059--1068.
\newblock \doi{10.1046/j.1365-8711.1998.01840.x}
\bibAnnoteFile{Andersson_1998_MNRAS}

\bibitem[{{Annala} et~al.(2020){Annala}, {Gorda}, {Kurkela}, {N{\"a}ttil{\"a}},
  and {Vuorinen}}]{2020NatPh..16..907A}
{Annala}, E., {Gorda}, T., {Kurkela}, A., {N{\"a}ttil{\"a}}, J., and
  {Vuorinen}, A. (2020).
\newblock {Evidence for quark-matter cores in massive neutron stars}.
\newblock \emph{Nature Physics} 16, 907--910.
\newblock \doi{10.1038/s41567-020-0914-9}
\bibAnnoteFile{2020NatPh..16..907A}

\bibitem[{{Bardeen} et~al.(1966){Bardeen}, {Thorne}, and
  {Meltzer}}]{1966ApJ...145..505B}
{Bardeen}, J.~M., {Thorne}, K.~S., and {Meltzer}, D.~W. (1966).
\newblock {A Catalogue of Methods for Studying the Normal Modes of Radial
  Pulsation of General-Relativistic Stellar Models}.
\newblock \emph{ApJ} 145, 505.
\newblock \doi{10.1086/148791}
\bibAnnoteFile{1966ApJ...145..505B}

\bibitem[{{Bauswein} et~al.(2010){Bauswein}, {Oechslin}, and
  {Janka}}]{Bauswein_2010_PhRvD}
{Bauswein}, A., {Oechslin}, R., and {Janka}, H.~T. (2010).
\newblock {Discriminating strange star mergers from neutron star mergers by
  gravitational-wave measurements}.
\newblock \emph{Phys. Rev. D.} 81, 024012.
\newblock \doi{10.1103/PhysRevD.81.024012}
\bibAnnoteFile{Bauswein_2010_PhRvD}

\bibitem[{{Benvenuto} and {Horvath}(1991)}]{Benvenuto_1991_MNRAS}
{Benvenuto}, O.~G. and {Horvath}, J.~E. (1991).
\newblock {Radial pulsations of strange stars and the internal composition of
  pulsars.}
\newblock \emph{Mon. Not. Roy. Astron. Soc.} 250, 679.
\newblock \doi{10.1093/mnras/250.4.679}
\bibAnnoteFile{Benvenuto_1991_MNRAS}

\bibitem[{{Berezhiani} et~al.(2003){Berezhiani}, {Bombaci}, {Drago},
  {Frontera}, and {Lavagno}}]{Berezhiani_2003_ApJ}
{Berezhiani}, Z., {Bombaci}, I., {Drago}, A., {Frontera}, F., and {Lavagno}, A.
  (2003).
\newblock {Gamma-Ray Bursts from Delayed Collapse of Neutron Stars to Quark
  Matter Stars}.
\newblock \emph{Astrophys. J.} 586, 1250--1253.
\newblock \doi{10.1086/367756}
\bibAnnoteFile{Berezhiani_2003_ApJ}

\bibitem[{{Bombaci} and {Datta}(2000)}]{Bombaci_2000_ApJ}
{Bombaci}, I. and {Datta}, B. (2000).
\newblock {Conversion of Neutron Stars to Strange Stars as the Central Engine
  of Gamma-Ray Bursts}.
\newblock \emph{Astrophys. J. Lett.} 530, L69--L72.
\newblock \doi{10.1086/312497}
\bibAnnoteFile{Bombaci_2000_ApJ}

\bibitem[{Bombaci et~al.(2021)Bombaci, Drago, Logoteta, Pagliara, and
  Vida\~na}]{I.Bombaci_2021_PRL}
Bombaci, I., Drago, A., Logoteta, D., Pagliara, G., and Vida\~na, I. (2021).
\newblock Was gw190814 a black hole--strange quark star system?
\newblock \emph{Phys. Rev. Lett.} 126, 162702.
\newblock \doi{10.1103/PhysRevLett.126.162702}
\bibAnnoteFile{I.Bombaci_2021_PRL}

\bibitem[{{Bora} and {Dev Goswami}(2021)}]{Bora_2021_MNRAS}
{Bora}, J. and {Dev Goswami}, U. (2021).
\newblock {Radial oscillations and gravitational wave echoes of strange stars
  for various equations of state}.
\newblock \emph{Mon. Not. Roy. Astron. Soc.} 502, 1557--1568.
\newblock \doi{10.1093/mnras/stab050}
\bibAnnoteFile{Bora_2021_MNRAS}

\bibitem[{{Buballa}(2005)}]{Buballa_2005_PhR}
{Buballa}, M. (2005).
\newblock {NJL-model analysis of dense quark matter [review article]}.
\newblock \emph{Phys. Rep.} 407, 205--376.
\newblock \doi{10.1016/j.physrep.2004.11.004}
\bibAnnoteFile{Buballa_2005_PhR}

\bibitem[{{Chandrasekhar}(1964{\natexlab{a}})}]{Chandrasekhar_1964_PhRvL}
{Chandrasekhar}, S. (1964{\natexlab{a}}).
\newblock {Dynamical Instability of Gaseous Masses Approaching the
  Schwarzschild Limit in General Relativity}.
\newblock \emph{Phys. Rev. Lett.} 12, 114--116.
\newblock \doi{10.1103/PhysRevLett.12.114}
\bibAnnoteFile{Chandrasekhar_1964_PhRvL}

\bibitem[{{Chandrasekhar}(1964{\natexlab{b}})}]{Chandrasekhar_1964_ApJ}
{Chandrasekhar}, S. (1964{\natexlab{b}}).
\newblock {The Dynamical Instability of Gaseous Masses Approaching the
  Schwarzschild Limit in General Relativity.}
\newblock \emph{Astrophys. J.} 140, 417.
\newblock \doi{10.1086/147938}
\bibAnnoteFile{Chandrasekhar_1964_ApJ}

\bibitem[{{Cheng} and {Dai}(1996)}]{Cheng_1996_PhRvL}
{Cheng}, K.~S. and {Dai}, Z.~G. (1996).
\newblock {Conversion of Neutron Stars to Strange Stars as a Possible Origin of
  {\ensuremath{\gamma}}-Ray Bursts}.
\newblock \emph{Phys. Rev. Lett.} 77, 1210--1213.
\newblock \doi{10.1103/PhysRevLett.77.1210}
\bibAnnoteFile{Cheng_1996_PhRvL}

\bibitem[{{Chodos} et~al.(1974{\natexlab{a}}){Chodos}, {Jaffe}, {Johnson}, and
  {Thorn}}]{Chodos_1974_1_PhRvD}
{Chodos}, A., {Jaffe}, R.~L., {Johnson}, K., and {Thorn}, C.~B.
  (1974{\natexlab{a}}).
\newblock {Baryon structure in the bag theory}.
\newblock \emph{Phys. Rev. D.} 10, 2599--2604.
\newblock \doi{10.1103/PhysRevD.10.2599}
\bibAnnoteFile{Chodos_1974_1_PhRvD}

\bibitem[{{Chodos} et~al.(1974{\natexlab{b}}){Chodos}, {Jaffe}, {Johnson},
  {Thorn}, and {Weisskopf}}]{Chodos_1974_PhRvD}
{Chodos}, A., {Jaffe}, R.~L., {Johnson}, K., {Thorn}, C.~B., and {Weisskopf},
  V.~F. (1974{\natexlab{b}}).
\newblock {New extended model of hadrons}.
\newblock \emph{Phys. Rev. D.} 9, 3471--3495.
\newblock \doi{10.1103/PhysRevD.9.3471}
\bibAnnoteFile{Chodos_1974_PhRvD}

\bibitem[{{Colpi} and {Miller}(1992)}]{Colpi_1992_ApJ}
{Colpi}, M. and {Miller}, J.~C. (1992).
\newblock {Rotational Properties of Strange Stars}.
\newblock \emph{Astrophys. J.} 388, 513.
\newblock \doi{10.1086/171170}
\bibAnnoteFile{Colpi_1992_ApJ}

\bibitem[{Cowling(1941)}]{Cowling_1941_MNRAS}
Cowling, T.~G. (1941).
\newblock {The Non-radial Oscillations of Polytropic Stars}.
\newblock \emph{Monthly Notices of the Royal Astronomical Society} 101,
  367--375.
\newblock \doi{10.1093/mnras/101.8.367}
\bibAnnoteFile{Cowling_1941_MNRAS}

\bibitem[{{Curi} et~al.(2022){Curi}, {Castro}, {Flores}, and
  {Lenzi}}]{Curi_2022_EPJC}
{Curi}, E. J.~A., {Castro}, L.~B., {Flores}, C.~V., and {Lenzi}, C.~H. (2022).
\newblock {Non-radial oscillations and global stellar properties of anisotropic
  compact stars using realistic equations of state}.
\newblock \emph{European Physical Journal C} 82, 527.
\newblock \doi{10.1140/epjc/s10052-022-10498-4}
\bibAnnoteFile{Curi_2022_EPJC}

\bibitem[{{Dai} et~al.(2016){Dai}, {Wang}, {Wu}, and {Huang}}]{Dai_2016_ApJ}
{Dai}, Z.~G., {Wang}, J.~S., {Wu}, X.~F., and {Huang}, Y.~F. (2016).
\newblock {Repeating Fast Radio Bursts from Highly Magnetized Pulsars Traveling
  through Asteroid Belts}.
\newblock \emph{Astrophys. J.} 829, 27.
\newblock \doi{10.3847/0004-637X/829/1/27}
\bibAnnoteFile{Dai_2016_ApJ}

\bibitem[{Damour and Nagar(2009)}]{Damour_2009_PRD}
Damour, T. and Nagar, A. (2009).
\newblock Relativistic tidal properties of neutron stars.
\newblock \emph{Phys. Rev. D} 80, 084035.
\newblock \doi{10.1103/PhysRevD.80.084035}
\bibAnnoteFile{Damour_2009_PRD}

\bibitem[{{Deb} et~al.(2017){Deb}, {Chowdhury}, {Ray}, {Rahaman}, and
  {Guha}}]{Deb_2017_AnPhy}
{Deb}, D., {Chowdhury}, S.~R., {Ray}, S., {Rahaman}, F., and {Guha}, B.~K.
  (2017).
\newblock {Relativistic model for anisotropic strange stars}.
\newblock \emph{Ann. Phys.} 387, 239--252.
\newblock \doi{10.1016/j.aop.2017.10.010}
\bibAnnoteFile{Deb_2017_AnPhy}

\bibitem[{{Di Clemente} et~al.(2023){Di Clemente}, {Drago}, {Char}, and
  {Pagliara}}]{DiClemente_2023_A&A}
{Di Clemente}, F., {Drago}, A., {Char}, P., and {Pagliara}, G. (2023).
\newblock {Stability and instability of strange dwarfs}.
\newblock \emph{Astron. Astrophys.} 678, L1.
\newblock \doi{10.1051/0004-6361/202347607}
\bibAnnoteFile{DiClemente_2023_A&A}

\bibitem[{{Drago} et~al.(2004){Drago}, {Lavagno}, and
  {Pagliara}}]{Drago_2004_EPJA}
{Drago}, A., {Lavagno}, A., and {Pagliara}, G. (2004).
\newblock {The Supernova-GRB connection}.
\newblock \emph{Eur. Phys. J. A.} 19, 197--201.
\newblock \doi{10.1140/epjad/s2004-03-033-9}
\bibAnnoteFile{Drago_2004_EPJA}

\bibitem[{Dutra et~al.(2014)Dutra, Louren\c{c}o, Avancini, Carlson, Delfino,
  Menezes et~al.}]{Dutra_2014_prc}
Dutra, M., Louren\c{c}o, O., Avancini, S.~S., Carlson, B.~V., Delfino, A.,
  Menezes, D.~P., et~al. (2014).
\newblock {Relativistic Mean-Field Hadronic Models under Nuclear Matter
  Constraints}.
\newblock \emph{Phys. Rev. C} 90, 055203.
\newblock \doi{10.1103/PhysRevC.90.055203}
\bibAnnoteFile{Dutra_2014_prc}

\bibitem[{{Eguchi} and {Sugawara}(1974)}]{Eguchi_1974_PhRvD}
{Eguchi}, T. and {Sugawara}, H. (1974).
\newblock {Extended model of elementary particles based on an analogy with
  superconductivity}.
\newblock \emph{Phys. Rev. D.} 10, 4257--4262.
\newblock \doi{10.1103/PhysRevD.10.4257}
\bibAnnoteFile{Eguchi_1974_PhRvD}

\bibitem[{{Einstein}(1916)}]{Einstein_1916_SPAW}
{Einstein}, A. (1916).
\newblock {N{\"a}herungsweise Integration der Feldgleichungen der Gravitation}.
\newblock \emph{Sitzungsber. Königl. Preuss. Akad. Wiss.} , 688--696
\bibAnnoteFile{Einstein_1916_SPAW}

\bibitem[{{Einstein}(1918)}]{Einstein_1918_SPAW}
{Einstein}, A. (1918).
\newblock {{\"U}ber Gravitationswellen}.
\newblock \emph{Sitzungsber. Königl. Preuss. Akad. Wiss.} , 154--167
\bibAnnoteFile{Einstein_1918_SPAW}

\bibitem[{Farhi and Jaffe(1984)}]{Farhi&Jaffe_1984_PhysRevD}
Farhi, E. and Jaffe, R.~L. (1984).
\newblock Strange matter.
\newblock \emph{Phys. Rev. D.} 30, 2379--2390.
\newblock \doi{10.1103/PhysRevD.30.2379}
\bibAnnoteFile{Farhi&Jaffe_1984_PhysRevD}

\bibitem[{{Flanagan} and {Hinderer}(2008)}]{Flanagan_2008_PhRvD}
{Flanagan}, {\'E}.~{\'E}. and {Hinderer}, T. (2008).
\newblock {Constraining neutron-star tidal Love numbers with gravitational-wave
  detectors}.
\newblock \emph{Phys. Rev. D.} 77, 021502.
\newblock \doi{10.1103/PhysRevD.77.021502}
\bibAnnoteFile{Flanagan_2008_PhRvD}

\bibitem[{Fraga et~al.(2001)Fraga, Pisarski, and
  Schaffner-Bielich}]{Fraga_2001_PRD}
Fraga, E.~S., Pisarski, R.~D., and Schaffner-Bielich, J. (2001).
\newblock {Small, dense quark stars from perturbative QCD}.
\newblock \emph{Phys. Rev. D} 63, 121702.
\newblock \doi{10.1103/PhysRevD.63.121702}
\bibAnnoteFile{Fraga_2001_PRD}

\bibitem[{{Geng} et~al.(2021){Geng}, {Li}, and {Huang}}]{Geng_2021_Innov}
{Geng}, J., {Li}, B., and {Huang}, Y. (2021).
\newblock {Repeating fast radio bursts from collapses of the crust of a strange
  star}.
\newblock \emph{Innov.} 2, 100152.
\newblock \doi{10.1016/j.xinn.2021.100152}
\bibAnnoteFile{Geng_2021_Innov}

\bibitem[{{Geng} and {Huang}(2015)}]{Geng_2015_ApJ}
{Geng}, J.~J. and {Huang}, Y.~F. (2015).
\newblock {Fast Radio Bursts: Collisions between Neutron Stars and
  Asteroids/Comets}.
\newblock \emph{Astrophys. J.} 809, 24.
\newblock \doi{10.1088/0004-637X/809/1/24}
\bibAnnoteFile{Geng_2015_ApJ}

\bibitem[{{Geng} et~al.(2015){Geng}, {Huang}, and {Lu}}]{Geng_2015_APJL}
{Geng}, J.~J., {Huang}, Y.~F., and {Lu}, T. (2015).
\newblock {Coalescence of Strange-quark Planets with Strange Stars: a New Kind
  of Source for Gravitational Wave Bursts}.
\newblock \emph{Astrophys. J.} 804, 21.
\newblock \doi{10.1088/0004-637X/804/1/21}
\bibAnnoteFile{Geng_2015_APJL}

\bibitem[{{Geng} et~al.(2020){Geng}, {Li}, {Li}, {Xiong}, {Kuiper}, and
  {Huang}}]{Geng_2020_ApJ}
{Geng}, J.-J., {Li}, B., {Li}, L.-B., {Xiong}, S.-L., {Kuiper}, R., and
  {Huang}, Y.-F. (2020).
\newblock {FRB 200428: An Impact between an Asteroid and a Magnetar}.
\newblock \emph{Astrophys. J. Lett.} 898, L55.
\newblock \doi{10.3847/2041-8213/aba83c}
\bibAnnoteFile{Geng_2020_ApJ}

\bibitem[{{Ghosh} et~al.(1995){Ghosh}, {Phatak}, and {Sahu}}]{Ghosh_1995_ZPhyA}
{Ghosh}, S.~K., {Phatak}, S.~C., and {Sahu}, P.~K. (1995).
\newblock {Hybrid stars and quark hadron phase transition in chiral colour
  dielectric model.}
\newblock \emph{Zeitschrift fur Physik A Hadrons and Nuclei} 352, 457--466.
\newblock \doi{10.1007/BF01299764}
\bibAnnoteFile{Ghosh_1995_ZPhyA}

\bibitem[{{Glendenning}(1996)}]{Glendenning_1996_book}
{Glendenning}, N. (1996).
\newblock \emph{{Compact Stars. Nuclear Physics, Particle Physics and General
  Relativity.}} (Springer).
\newblock \doi{10.1007/978-1-4684-0491-3}
\bibAnnoteFile{Glendenning_1996_book}

\bibitem[{{Glendenning}(1992)}]{Glendenning_1992_PhRvD}
{Glendenning}, N.~K. (1992).
\newblock {First-order phase transitions with more than one conserved charge:
  Consequences for neutron stars}.
\newblock \emph{Phys. Rev. D} 46, 1274--1287.
\newblock \doi{10.1103/PhysRevD.46.1274}
\bibAnnoteFile{Glendenning_1992_PhRvD}

\bibitem[{{Glendenning} et~al.(1995){Glendenning}, {Kettner}, and
  {Weber}}]{Glendenning_1995_ApJ}
{Glendenning}, N.~K., {Kettner}, C., and {Weber}, F. (1995).
\newblock {From Strange Stars to Strange Dwarfs}.
\newblock \emph{Astrophys. J.} 450, 253.
\newblock \doi{10.1086/176136}
\bibAnnoteFile{Glendenning_1995_ApJ}

\bibitem[{{Glendenning} and {Weber}(1992)}]{Glendenning_1992_ApJ}
{Glendenning}, N.~K. and {Weber}, F. (1992).
\newblock {Nuclear Solid Crust on Rotating Strange Quark Stars}.
\newblock \emph{Astrophys. J.} 400, 647.
\newblock \doi{10.1086/172026}
\bibAnnoteFile{Glendenning_1992_ApJ}

\bibitem[{{Gon{\c{c}}alves} et~al.(2023){Gon{\c{c}}alves}, {Jim{\'e}nez}, and
  {Lazzari}}]{Goncalves_2023_EPJA}
{Gon{\c{c}}alves}, V.~P., {Jim{\'e}nez}, J.~C., and {Lazzari}, L. (2023).
\newblock {Revisiting the stability of strange-dwarf stars and strange
  planets}.
\newblock \emph{Eur. Phys. J. A.} 59, 251.
\newblock \doi{10.1140/epja/s10050-023-01175-5}
\bibAnnoteFile{Goncalves_2023_EPJA}

\bibitem[{{Gondek-Rosi{\'n}ska} et~al.(2003){Gondek-Rosi{\'n}ska},
  {Gourgoulhon}, and {Haensel}}]{Gondek_2003_A&A}
{Gondek-Rosi{\'n}ska}, D., {Gourgoulhon}, E., and {Haensel}, P. (2003).
\newblock {Are rotating strange quark stars good sources of gravitational
  waves?}
\newblock \emph{Astron. Astrophys.} 412, 777--790.
\newblock \doi{10.1051/0004-6361:20031431}
\bibAnnoteFile{Gondek_2003_A&A}

\bibitem[{{Gondek-Rosinska} and {Limousin}(2008)}]{Gondek-Rosinska_2008_arXiv}
{Gondek-Rosinska}, D. and {Limousin}, F. (2008).
\newblock {The final phase of inspiral of strange quark star binaries}.
\newblock \emph{arXiv e-prints} , arXiv:0801.4829\doi{10.48550/arXiv.0801.4829}
\bibAnnoteFile{Gondek-Rosinska_2008_arXiv}

\bibitem[{{Gorenstein} and {Yang}(1995)}]{Gorenstein_1995_PhRvD}
{Gorenstein}, M.~I. and {Yang}, S.~N. (1995).
\newblock {Gluon plasma with a medium-dependent dispersion relation}.
\newblock \emph{Phys. Rev. D.} 52, 5206--5212.
\newblock \doi{10.1103/PhysRevD.52.5206}
\bibAnnoteFile{Gorenstein_1995_PhRvD}

\bibitem[{Gusakov et~al.(2005)Gusakov, Kaminker, Yakovlev, and
  Gnedin}]{Gusakov_2005_MNRAS}
Gusakov, M.~E., Kaminker, A.~D., Yakovlev, D.~G., and Gnedin, O.~Y. (2005).
\newblock {Cooling of Akmal-Pandharipande-Ravenhall neutron star models}.
\newblock \emph{Mon. Not. Roy. Astron. Soc.} 363, 555--562.
\newblock \doi{10.1111/j.1365-2966.2005.09459.x}
\bibAnnoteFile{Gusakov_2005_MNRAS}

\bibitem[{{Harry} and {LIGO Scientific Collaboration}(2010)}]{Harry_2010}
{Harry}, G.~M. and {LIGO Scientific Collaboration} (2010).
\newblock {Advanced LIGO: the next generation of gravitational wave detectors}.
\newblock \emph{Class. Quantum Gravity.} 27, 084006.
\newblock \doi{10.1088/0264-9381/27/8/084006}
\bibAnnoteFile{Harry_2010}

\bibitem[{{He} et~al.(2007){He}, {Sun}, {Feng}, and {Zong}}]{He_2007_JPhG}
{He}, M., {Sun}, W.-m., {Feng}, H.-t., and {Zong}, H.-s. (2007).
\newblock {A model study of QCD phase transition}.
\newblock \emph{J. Phys. G: Nucl. Part. Phys.} 34, 2655--2663.
\newblock \doi{10.1088/0954-3899/34/12/010}
\bibAnnoteFile{He_2007_JPhG}

\bibitem[{{Hild} et~al.(2008){Hild}, {Chelkowski}, and
  {Freise}}]{Hild_2008arXiv}
{Hild}, S., {Chelkowski}, S., and {Freise}, A. (2008).
\newblock {Pushing towards the ET sensitivity using `conventional' technology}.
\newblock \emph{arXiv e-prints} , arXiv:0810.0604\doi{10.48550/arXiv.0810.0604}
\bibAnnoteFile{Hild_2008arXiv}

\bibitem[{Hinderer(2008)}]{Hinderer_2008_apj}
Hinderer, T. (2008).
\newblock {Tidal Love numbers of neutron stars}.
\newblock \emph{Astrophys. J.} 677, 1216--1220.
\newblock \doi{10.1086/533487}.
\newblock [Erratum: Astrophys.J. 697, 964 (2009)]
\bibAnnoteFile{Hinderer_2008_apj}

\bibitem[{{Hu} and {Huang}(2023)}]{Hu_2023_ApJS}
{Hu}, C.-R. and {Huang}, Y.-F. (2023).
\newblock {A Comprehensive Analysis of Repeating Fast Radio Bursts}.
\newblock \emph{Astrophys. J. Suppl. Ser.} 269, 17.
\newblock \doi{10.3847/1538-4365/acf566}
\bibAnnoteFile{Hu_2023_ApJS}

\bibitem[{{Huang} and {Lu}(1997)}]{Huang_1997_A&A}
{Huang}, Y.~F. and {Lu}, T. (1997).
\newblock {Strange stars: how dense can their crust be?}
\newblock \emph{A\&A} 325, 189--194
\bibAnnoteFile{Huang_1997_A&A}

\bibitem[{{Huang} and {Yu}(2017)}]{Huang_2017_ApJ}
{Huang}, Y.~F. and {Yu}, Y.~B. (2017).
\newblock {Searching for Strange Quark Matter Objects in Exoplanets}.
\newblock \emph{Astrophys. J.} 848, 115.
\newblock \doi{10.3847/1538-4357/aa8b63}
\bibAnnoteFile{Huang_2017_ApJ}

\bibitem[{{Ivanenko} and {Kurdgelaidze}(1965)}]{1965Ap......1..251I}
{Ivanenko}, D.~D. and {Kurdgelaidze}, D.~F. (1965).
\newblock {Hypothesis concerning quark stars}.
\newblock \emph{Astrophysics} 1, 251--252.
\newblock \doi{10.1007/BF01042830}
\bibAnnoteFile{1965Ap......1..251I}

\bibitem[{{Jia} and {Huang}(2004)}]{Jia_2004_ChA&A}
{Jia}, J.-j. and {Huang}, Y.-f. (2004).
\newblock {A numerical study of the collapse of the crust of strange stars}.
\newblock \emph{Chin. Astron. Astrophys.} 28, 144--153.
\newblock \doi{10.1016/S0275-1062(04)90017-3}
\bibAnnoteFile{Jia_2004_ChA&A}

\bibitem[{Jim\'enez and Fraga(2019)}]{Jimenez_2019_PRD}
Jim\'enez, J.~C. and Fraga, E.~S. (2019).
\newblock {Radial oscillations of quark stars from perturbative QCD}.
\newblock \emph{Phys. Rev. D} 100, 114041.
\newblock \doi{10.1103/PhysRevD.100.114041}
\bibAnnoteFile{Jimenez_2019_PRD}

\bibitem[{{Kikkawa}(1976)}]{Kikkawa_1976_PThPh}
{Kikkawa}, K. (1976).
\newblock {Quantum Corrections in Superconductor Models}.
\newblock \emph{Prog. Theor. Phys.} 56, 947--955.
\newblock \doi{10.1143/PTP.56.947}
\bibAnnoteFile{Kikkawa_1976_PThPh}

\bibitem[{{Klevansky}(1992)}]{Klevansky_1992_RvMP}
{Klevansky}, S.~P. (1992).
\newblock {The Nambu-Jona-Lasinio model of quantum chromodynamics}.
\newblock \emph{Rev. Mod. Phys.} 64, 649--708.
\newblock \doi{10.1103/RevModPhys.64.649}
\bibAnnoteFile{Klevansky_1992_RvMP}

\bibitem[{{Kokkotas} and {Schutz}(1992)}]{Kokkotas_1992_MNRAS}
{Kokkotas}, K.~D. and {Schutz}, B.~F. (1992).
\newblock {W-modes - A new family of normal modes of pulsating relativistic
  stars}.
\newblock \emph{Monthly Notices of the Royal Astronomical Society} 255,
  119--128.
\newblock \doi{10.1093/mnras/255.1.119}
\bibAnnoteFile{Kokkotas_1992_MNRAS}

\bibitem[{{Kuerban} et~al.(2020){Kuerban}, {Geng}, {Huang}, {Zong}, and
  {Gong}}]{Kuerban_2020_ApJ}
{Kuerban}, A., {Geng}, J.-J., {Huang}, Y.-F., {Zong}, H.-S., and {Gong}, H.
  (2020).
\newblock {Close-in Exoplanets as Candidates for Strange Quark Matter Objects}.
\newblock \emph{Astrophys. J.} 890, 41.
\newblock \doi{10.3847/1538-4357/ab698b}
\bibAnnoteFile{Kuerban_2020_ApJ}

\bibitem[{{Kumar} et~al.(2022){Kumar}, {Thapa}, and
  {Sinha}}]{AnilKumar_2022_MN}
{Kumar}, A., {Thapa}, V.~B., and {Sinha}, M. (2022).
\newblock {Compact star merger events with stars composed of interacting
  strange quark matter}.
\newblock \emph{Mon. Not. R. Astron. Soc.} 513, 3788--3797.
\newblock \doi{10.1093/mnras/stac1150}
\bibAnnoteFile{AnilKumar_2022_MN}

\bibitem[{{Kurban} et~al.(2022{\natexlab{a}}){Kurban}, {Huang}, {Geng}, {Li},
  {Xu}, {Wang} et~al.}]{Kurban_2022_ApJ}
{Kurban}, A., {Huang}, Y.-F., {Geng}, J.-J., {Li}, B., {Xu}, F., {Wang}, X.,
  et~al. (2022{\natexlab{a}}).
\newblock {Periodic Repeating Fast Radio Bursts: Interaction between a
  Magnetized Neutron Star and Its Planet in an Eccentric Orbit}.
\newblock \emph{Astrophys. J.} 928, 94.
\newblock \doi{10.3847/1538-4357/ac558f}
\bibAnnoteFile{Kurban_2022_ApJ}

\bibitem[{{Kurban} et~al.(2022{\natexlab{b}}){Kurban}, {Huang}, {Geng}, and
  {Zong}}]{Kurban_2022_PhLB}
{Kurban}, A., {Huang}, Y.-F., {Geng}, J.-J., and {Zong}, H.-S.
  (2022{\natexlab{b}}).
\newblock {Searching for strange quark matter objects among white dwarfs}.
\newblock \emph{Phys. Lett. B.} 832, 137204.
\newblock \doi{10.1016/j.physletb.2022.137204}
\bibAnnoteFile{Kurban_2022_PhLB}

\bibitem[{{Levan} et~al.(2016){Levan}, {Crowther}, {de Grijs}, {Langer}, {Xu},
  and {Yoon}}]{Levan_2016_SSR}
{Levan}, A., {Crowther}, P., {de Grijs}, R., {Langer}, N., {Xu}, D., and
  {Yoon}, S.-C. (2016).
\newblock {Gamma-Ray Burst Progenitors}.
\newblock \emph{Space Science Reviews} 202, 33--78.
\newblock \doi{10.1007/s11214-016-0312-x}
\bibAnnoteFile{Levan_2016_SSR}

\bibitem[{Li et~al.(2010)Li, Zhou, Burgio, and Schulze}]{Li_2010_PhRvC}
Li, A., Zhou, X.~R., Burgio, G.~F., and Schulze, H.~J. (2010).
\newblock {Protoneutron stars in the Brueckner-Hartree-Fock approach and
  finite-temperature kaon condensation}.
\newblock \emph{Phys. Rev. C} 81, 025806.
\newblock \doi{10.1103/PhysRevC.81.025806}
\bibAnnoteFile{Li_2010_PhRvC}

\bibitem[{{Li} et~al.(2017){Li}, {Huang}, {Zhang}, {Li}, and
  {Li}}]{Li_2017_RAA}
{Li}, L.-B., {Huang}, Y.-F., {Zhang}, Z.-B., {Li}, D., and {Li}, B. (2017).
\newblock {Intensity distribution function and statistical properties of fast
  radio bursts}.
\newblock \emph{Res. Astron. Astrophys.} 17, 6.
\newblock \doi{10.1088/1674-4527/17/1/6}
\bibAnnoteFile{Li_2017_RAA}

\bibitem[{{Li} et~al.(2021){Li}, {Dong}, {Zhang}, and {Li}}]{Li_2021_ApJ}
{Li}, X.~J., {Dong}, X.~F., {Zhang}, Z.~B., and {Li}, D. (2021).
\newblock {Long and Short Fast Radio Bursts Are Different from Repeating and
  Nonrepeating Transients}.
\newblock \emph{Astrophys. J.} 923, 230.
\newblock \doi{10.3847/1538-4357/ac3085}
\bibAnnoteFile{Li_2021_ApJ}

\bibitem[{{Li} and {Schulze}(2012)}]{Li_2012_PhRvC}
{Li}, Z.~H. and {Schulze}, H.~J. (2012).
\newblock {Nuclear matter with chiral forces in Brueckner-Hartree-Fock
  approximation}.
\newblock \emph{Phys. Rev. C} 85, 064002.
\newblock \doi{10.1103/PhysRevC.85.064002}
\bibAnnoteFile{Li_2012_PhRvC}

\bibitem[{{Limousin} et~al.(2005){Limousin}, {Gondek-Rosi{\'n}ska}, and
  {Gourgoulhon}}]{Limousin_2005_PhRvD}
{Limousin}, F., {Gondek-Rosi{\'n}ska}, D., and {Gourgoulhon}, E. (2005).
\newblock {Last orbits of binary strange quark stars}.
\newblock \emph{Phys. Rev. D.} 71, 064012.
\newblock \doi{10.1103/PhysRevD.71.064012}
\bibAnnoteFile{Limousin_2005_PhRvD}

\bibitem[{{Lin} et~al.(2006){Lin}, {Cheng}, {Chu}, and {Suen}}]{Lin_2006_ApJ}
{Lin}, L.~M., {Cheng}, K.~S., {Chu}, M.~C., and {Suen}, W.~M. (2006).
\newblock {Gravitational Waves from Phase-Transition-Induced Collapse of
  Neutron Stars}.
\newblock \emph{Astrophys. J.} 639, 382--396.
\newblock \doi{10.1086/499202}
\bibAnnoteFile{Lin_2006_ApJ}

\bibitem[{{Lohakare} et~al.(2023){Lohakare}, {Maurya}, {Singh}, {Mishra}, and
  {Errehymy}}]{Lohakare_2023_MNRAS}
{Lohakare}, S.~V., {Maurya}, S.~K., {Singh}, K.~N., {Mishra}, B., and
  {Errehymy}, A. (2023).
\newblock {Influence of three parameters on maximum mass and stability of
  strange star under linear f(Q) - action}.
\newblock \emph{Mon. Not. R. Astron. Soc.} 526, 3796--3814.
\newblock \doi{10.1093/mnras/stad2861}
\bibAnnoteFile{Lohakare_2023_MNRAS}

\bibitem[{{Lopes} and {Menezes}(2022)}]{Lopes_2022_ApJ}
{Lopes}, L.~L. and {Menezes}, D.~P. (2022).
\newblock {On the Nature of the Mass-gap Object in the GW190814 Event}.
\newblock \emph{Astrophys. J.} 936, 41.
\newblock \doi{10.3847/1538-4357/ac81c4}
\bibAnnoteFile{Lopes_2022_ApJ}

\bibitem[{{Lorimer} et~al.(2007){Lorimer}, {Bailes}, {McLaughlin}, {Narkevic},
  and {Crawford}}]{Lorimer_2007_Science}
{Lorimer}, D.~R., {Bailes}, M., {McLaughlin}, M.~A., {Narkevic}, D.~J., and
  {Crawford}, F. (2007).
\newblock {A Bright Millisecond Radio Burst of Extragalactic Origin}.
\newblock \emph{Sci.} 318, 777.
\newblock \doi{10.1126/science.1147532}
\bibAnnoteFile{Lorimer_2007_Science}

\bibitem[{{L{\"u}} et~al.(2009){L{\"u}}, {Wu}, and {Zeng}}]{Lv_2009_RAA}
{L{\"u}}, Z.-K., {Wu}, S.-W., and {Zeng}, Z.-C. (2009).
\newblock {Gravitational wave radiation from a double white dwarf system inside
  our galaxy: a potential method for seeking strange dwarfs}.
\newblock \emph{Res. Astron. Astrophys.} 9, 745--750.
\newblock \doi{10.1088/1674-4527/9/7/002}
\bibAnnoteFile{Lv_2009_RAA}

\bibitem[{{Mallick} and {Sahu}(2014)}]{Mallick_2014_NuPhA}
{Mallick}, R. and {Sahu}, P.~K. (2014).
\newblock {Phase transitions in neutron star and magnetars and their connection
  with high energetic bursts in astrophysics}.
\newblock \emph{Nucl. Phys. A.} 921, 96--113.
\newblock \doi{10.1016/j.nuclphysa.2013.11.009}
\bibAnnoteFile{Mallick_2014_NuPhA}

\bibitem[{Matsuoka et~al.(2018)Matsuoka, Tsue, Da~Provid\^encia, Provid\^encia,
  and Yamamura}]{Matsuoka_2018_prd}
Matsuoka, H., Tsue, Y., Da~Provid\^encia, J.~a., Provid\^encia, C., and
  Yamamura, M. (2018).
\newblock {Hybrid stars from the NJL model with a tensor-interaction}.
\newblock \emph{Phys. Rev. D} 98, 074027.
\newblock \doi{10.1103/PhysRevD.98.074027}
\bibAnnoteFile{Matsuoka_2018_prd}

\bibitem[{{McDermott} et~al.(1988){McDermott}, {van Horn}, and
  {Hansen}}]{McDermott_1988_ApJ}
{McDermott}, P.~N., {van Horn}, H.~M., and {Hansen}, C.~J. (1988).
\newblock {Nonradial Oscillations of Neutron Stars}.
\newblock \emph{Astrophys. J.} 325, 725.
\newblock \doi{10.1086/166044}
\bibAnnoteFile{McDermott_1988_ApJ}

\bibitem[{{Menezes}(2021)}]{2021Univ....7..267M}
{Menezes}, D.~P. (2021).
\newblock {A Neutron Star Is Born}.
\newblock \emph{Universe} 7, 267.
\newblock \doi{10.3390/universe7080267}
\bibAnnoteFile{2021Univ....7..267M}

\bibitem[{{Miao} et~al.(2021){Miao}, {Jiang}, {Li}, and
  {Chen}}]{Miao_2021_ApJL}
{Miao}, Z., {Jiang}, J.-L., {Li}, A., and {Chen}, L.-W. (2021).
\newblock {Bayesian Inference of Strange Star Equation of State Using the
  GW170817 and GW190425 Data}.
\newblock \emph{Astrophys. J. Lett.} 917, L22.
\newblock \doi{10.3847/2041-8213/ac194d}
\bibAnnoteFile{Miao_2021_ApJL}

\bibitem[{{Moraes} and {Miranda}(2014)}]{Moraes_2014_MNRAS}
{Moraes}, P. H.~R.~S. and {Miranda}, O.~D. (2014).
\newblock {Probing strange stars with advanced gravitational wave detectors}.
\newblock \emph{Mon. Not. R. Astron. Soc.} 445, L11--L15.
\newblock \doi{10.1093/mnrasl/slu124}
\bibAnnoteFile{Moraes_2014_MNRAS}

\bibitem[{{Nambu} and {Jona-Lasinio}(1961{\natexlab{a}})}]{Nambu_1961_PhRv}
{Nambu}, Y. and {Jona-Lasinio}, G. (1961{\natexlab{a}}).
\newblock {Dynamical Model of Elementary Particles Based on an Analogy with
  Superconductivity. I}.
\newblock \emph{Phys. Rev.} 122, 345--358.
\newblock \doi{10.1103/PhysRev.122.345}
\bibAnnoteFile{Nambu_1961_PhRv}

\bibitem[{{Nambu} and {Jona-Lasinio}(1961{\natexlab{b}})}]{Nambu_1961_1_PhRv}
{Nambu}, Y. and {Jona-Lasinio}, G. (1961{\natexlab{b}}).
\newblock {Dynamical Model of Elementary Particles Based on an Analogy with
  Superconductivity. II}.
\newblock \emph{Phys. Rev.} 124, 246--254.
\newblock \doi{10.1103/PhysRev.124.246}
\bibAnnoteFile{Nambu_1961_1_PhRv}

\bibitem[{{Nurmamat} et~al.(2024){Nurmamat}, {Huang}, {Geng}, {Kurban}, and
  {Li}}]{Nurmamat_2024_EPJC}
{Nurmamat}, N., {Huang}, Y.-F., {Geng}, J.-J., {Kurban}, A., and {Li}, B.
  (2024).
\newblock {Repeating fast radio bursts produced by a strange star interacting
  with its planet in an eccentric orbit}.
\newblock \emph{European Physical Journal C} 84, 210.
\newblock \doi{10.1140/epjc/s10052-024-12572-5}
\bibAnnoteFile{Nurmamat_2024_EPJC}

\bibitem[{Oikonomou and Moustakidis(2023)}]{Oikonomou_2023_PhysRevD}
Oikonomou, P.~T. and Moustakidis, C.~C. (2023).
\newblock Color-flavor locked quark stars in light of the compact object in the
  hess j1731-347 and the gw190814 event.
\newblock \emph{Phys. Rev. D.} 108, 063010.
\newblock \doi{10.1103/PhysRevD.108.063010}
\bibAnnoteFile{Oikonomou_2023_PhysRevD}

\bibitem[{{Oppenheimer} and {Volkoff}(1939)}]{Oppenheimer_1939_PhRv}
{Oppenheimer}, J.~R. and {Volkoff}, G.~M. (1939).
\newblock {On Massive Neutron Cores}.
\newblock \emph{Phys. Rev.} 55, 374--381.
\newblock \doi{10.1103/PhysRev.55.374}
\bibAnnoteFile{Oppenheimer_1939_PhRv}

\bibitem[{{Perot} et~al.(2023){Perot}, {Chamel}, and
  {Vallet}}]{Perot_2023_PhRvD}
{Perot}, L., {Chamel}, N., and {Vallet}, P. (2023).
\newblock {Unmasking strange dwarfs with gravitational-wave observations}.
\newblock \emph{Phys. Rev. D.} 107, 103004.
\newblock \doi{10.1103/PhysRevD.107.103004}
\bibAnnoteFile{Perot_2023_PhRvD}

\bibitem[{{Peshier} et~al.(1994){Peshier}, {K{\"a}mpfer}, {Pavlenko}, and
  {Soff}}]{Peshier_1994_PhLB}
{Peshier}, A., {K{\"a}mpfer}, B., {Pavlenko}, O.~P., and {Soff}, G. (1994).
\newblock {An effective model of the quark-gluon plasma with thermal parton
  masses}.
\newblock \emph{Phys. Lett. B.} 337, 235--239.
\newblock \doi{10.1016/0370-2693(94)90969-5}
\bibAnnoteFile{Peshier_1994_PhLB}

\bibitem[{Peshier et~al.(2000)Peshier, Kampfer, and Soff}]{Peshier_1999_PRC}
Peshier, A., Kampfer, B., and Soff, G. (2000).
\newblock {The Equation of state of deconfined matter at finite chemical
  potential in a quasiparticle description}.
\newblock \emph{Phys. Rev. C} 61, 045203.
\newblock \doi{10.1103/PhysRevC.61.045203}
\bibAnnoteFile{Peshier_1999_PRC}

\bibitem[{{Peshier} et~al.(2002){Peshier}, {K{\"a}mpfer}, and
  {Soff}}]{Peshier_2002_PhRvD}
{Peshier}, A., {K{\"a}mpfer}, B., and {Soff}, G. (2002).
\newblock {From QCD lattice calculations to the equation of state of quark
  matter}.
\newblock \emph{Phys. Rev. D.} 66, 094003.
\newblock \doi{10.1103/PhysRevD.66.094003}
\bibAnnoteFile{Peshier_2002_PhRvD}

\bibitem[{Postnikov et~al.(2010)Postnikov, Prakash, and
  Lattimer}]{Postnikov_2010_PRD}
Postnikov, S., Prakash, M., and Lattimer, J.~M. (2010).
\newblock Tidal love numbers of neutron and self-bound quark stars.
\newblock \emph{Phys. Rev. D} 82, 024016.
\newblock \doi{10.1103/PhysRevD.82.024016}
\bibAnnoteFile{Postnikov_2010_PRD}

\bibitem[{{Prasad} and {Mallick}(2018)}]{Prasad_2018_ApJ}
{Prasad}, R. and {Mallick}, R. (2018).
\newblock {Dynamical Phase Transition in Neutron Stars}.
\newblock \emph{Astrophys. J.} 859, 57.
\newblock \doi{10.3847/1538-4357/aabf3b}
\bibAnnoteFile{Prasad_2018_ApJ}

\bibitem[{{Price} and {Thorne}(1969)}]{Price_1969_ApJ}
{Price}, R. and {Thorne}, K.~S. (1969).
\newblock {Non-Radial Pulsation of General-Relativistic Stellar Models. II.
  Properties of the Gravitational Waves}.
\newblock \emph{Astrophys. J.} 155, 163.
\newblock \doi{10.1086/149857}
\bibAnnoteFile{Price_1969_ApJ}

\bibitem[{{Rather} et~al.(2023){Rather}, {Panotopoulos}, and
  {Lopes}}]{Rather_2023_EPJC}
{Rather}, I.~A., {Panotopoulos}, G., and {Lopes}, I. (2023).
\newblock {Quark models and radial oscillations: decoding the HESS J1731-347
  compact object's equation of state}.
\newblock \emph{European Physical Journal C} 83, 1065.
\newblock \doi{10.1140/epjc/s10052-023-12223-1}
\bibAnnoteFile{Rather_2023_EPJC}

\bibitem[{Rebhan and Romatschke(2003)}]{Rebhan_2003_PhRvD}
Rebhan, A. and Romatschke, P. (2003).
\newblock {HTL quasiparticle models of deconfined QCD at finite chemical
  potential}.
\newblock \emph{Phys. Rev. D} 68, 025022.
\newblock \doi{10.1103/PhysRevD.68.025022}
\bibAnnoteFile{Rebhan_2003_PhRvD}

\bibitem[{Roupas et~al.(2021)Roupas, Panotopoulos, and
  Lopes}]{ZachariasRoupas_2021_PRD}
Roupas, Z., Panotopoulos, G., and Lopes, I. (2021).
\newblock Qcd color superconductivity in compact stars: Color-flavor locked
  quark star candidate for the gravitational-wave signal gw190814.
\newblock \emph{Phys. Rev. D.} 103, 083015.
\newblock \doi{10.1103/PhysRevD.103.083015}
\bibAnnoteFile{ZachariasRoupas_2021_PRD}

\bibitem[{{Rupak} and {Jaikumar}(2013)}]{Rupak_2013_PhRvC}
{Rupak}, G. and {Jaikumar}, P. (2013).
\newblock {r-mode instability in quark stars with a crystalline crust}.
\newblock \emph{Phys. Rev. C.} 88, 065801.
\newblock \doi{10.1103/PhysRevC.88.065801}
\bibAnnoteFile{Rupak_2013_PhRvC}

\bibitem[{Sagun et~al.(2023)Sagun, Giangrandi, Dietrich, Ivanytskyi, Negreiros,
  and Providência}]{Sagun_2023_ApJ}
Sagun, V., Giangrandi, E., Dietrich, T., Ivanytskyi, O., Negreiros, R., and
  Providência, C. (2023).
\newblock What is the nature of the hess j1731-347 compact object?
\newblock \emph{Astrophys. J.} 958, 49.
\newblock \doi{10.3847/1538-4357/acfc9e}
\bibAnnoteFile{Sagun_2023_ApJ}

\bibitem[{{Schertler} et~al.(1998){Schertler}, {Greiner}, {Sahu}, and
  {Thoma}}]{Schertler_1998_NuPhA}
{Schertler}, K., {Greiner}, C., {Sahu}, P.~K., and {Thoma}, M.~H. (1998).
\newblock {The influence of medium effects on the gross structure of hybrid
  stars}.
\newblock \emph{Nucl. Phys. A} 637, 451--465.
\newblock \doi{10.1016/S0375-9474(98)00330-3}
\bibAnnoteFile{Schertler_1998_NuPhA}

\bibitem[{{Schertler} et~al.(2000){Schertler}, {Greiner}, {Schaffner-Bielich},
  and {Thoma2}}]{Schertler_2000_NuPhA}
{Schertler}, K., {Greiner}, C., {Schaffner-Bielich}, J., and {Thoma2}, M.~H.
  (2000).
\newblock {Quark phases in neutron stars and a third family of compact stars as
  signature for phase transitions$^{1}$}.
\newblock \emph{Nuc. Phys. A} 677, 463--490.
\newblock \doi{10.1016/S0375-9474(00)00305-5}
\bibAnnoteFile{Schertler_2000_NuPhA}

\bibitem[{{Schertler} et~al.(1997){Schertler}, {Greiner}, and
  {Thoma}}]{Schertler_1997_NuPhA}
{Schertler}, K., {Greiner}, C., and {Thoma}, M.~H. (1997).
\newblock {Medium effects in strange quark matter and strange stars}.
\newblock \emph{Nucl. Phys. A} 616, 659--679.
\newblock \doi{10.1016/S0375-9474(97)00014-6}
\bibAnnoteFile{Schertler_1997_NuPhA}

\bibitem[{Schneider et~al.(2019)Schneider, Constantinou, Muccioli, and
  Prakash}]{Schneider_2019_prc}
Schneider, A.~S., Constantinou, C., Muccioli, B., and Prakash, M. (2019).
\newblock {Akmal-Pandharipande-Ravenhall equation of state for simulations of
  supernovae, neutron stars, and binary mergers}.
\newblock \emph{Phys. Rev. C} 100, 025803.
\newblock \doi{10.1103/PhysRevC.100.025803}
\bibAnnoteFile{Schneider_2019_prc}

\bibitem[{{Shafeeque} et~al.(2023){Shafeeque}, {Mathew}, and
  {Nandy}}]{Shafeeque_2023_JApA}
{Shafeeque}, M., {Mathew}, A., and {Nandy}, M.~K. (2023).
\newblock {Maximal mass of the neutron star with a deconfined quark core}.
\newblock \emph{J. Astrophys. Astron.} 44, 63.
\newblock \doi{10.1007/s12036-023-09957-5}
\bibAnnoteFile{Shafeeque_2023_JApA}

\bibitem[{{Shu} et~al.(2017){Shu}, {Huang}, and {Zong}}]{Shu_2017_MPLA}
{Shu}, X.-Y., {Huang}, Y.-F., and {Zong}, H.-S. (2017).
\newblock {Gamma-ray bursts generated from phase transition of neutron stars to
  quark stars}.
\newblock \emph{Mod. Phys. Lett. A.} 32, 1750027.
\newblock \doi{10.1142/S0217732317500274}
\bibAnnoteFile{Shu_2017_MPLA}

\bibitem[{Sotani et~al.(2011)Sotani, Yasutake, Maruyama, and
  Tatsumi}]{Sotani_PRD_2011}
Sotani, H., Yasutake, N., Maruyama, T., and Tatsumi, T. (2011).
\newblock Signatures of hadron-quark mixed phase in gravitational waves.
\newblock \emph{Phys. Rev. D} 83, 024014.
\newblock \doi{10.1103/PhysRevD.83.024014}
\bibAnnoteFile{Sotani_PRD_2011}

\bibitem[{Tak\'atsy and Kov\'acs(2020)}]{Tak_2020_PRD}
Tak\'atsy, J. and Kov\'acs, P. (2020).
\newblock Comment on ``tidal love numbers of neutron and self-bound quark
  stars''.
\newblock \emph{Phys. Rev. D} 102, 028501.
\newblock \doi{10.1103/PhysRevD.102.028501}
\bibAnnoteFile{Tak_2020_PRD}

\bibitem[{{Thorne}(1998)}]{Thorne_1998_PhRvD}
{Thorne}, K.~S. (1998).
\newblock {Tidal stabilization of rigidly rotating, fully relativistic neutron
  stars}.
\newblock \emph{Phys. Rev. D.} 58, 124031.
\newblock \doi{10.1103/PhysRevD.58.124031}
\bibAnnoteFile{Thorne_1998_PhRvD}

\bibitem[{{Thorne} and {Campolattaro}(1967)}]{Thorne_1967_ApJ}
{Thorne}, K.~S. and {Campolattaro}, A. (1967).
\newblock {Non-Radial Pulsation of General-Relativistic Stellar Models. I.
  Analytic Analysis for L \textgreater= 2}.
\newblock \emph{Astrophys. J.} 149, 591.
\newblock \doi{10.1086/149288}
\bibAnnoteFile{Thorne_1967_ApJ}

\bibitem[{{Thornton} et~al.(2013){Thornton}, {Stappers}, {Bailes}, {Barsdell},
  {Bates}, {Bhat} et~al.}]{Thornton_2013_Science}
{Thornton}, D., {Stappers}, B., {Bailes}, M., {Barsdell}, B., {Bates}, S.,
  {Bhat}, N.~D.~R., et~al. (2013).
\newblock {A Population of Fast Radio Bursts at Cosmological Distances}.
\newblock \emph{Sci.} 341, 53--56.
\newblock \doi{10.1126/science.1236789}
\bibAnnoteFile{Thornton_2013_Science}

\bibitem[{{Tolman}(1934)}]{Tolman_1934_rtc_book}
{Tolman}, R.~C. (1934).
\newblock \emph{{Relativity, Thermodynamics, and Cosmology}}, vol. 497.
\newblock \doi{10.1126/science.80.2077.358}
\bibAnnoteFile{Tolman_1934_rtc_book}

\bibitem[{{Tolman}(1939)}]{Tolman_1939_PhRev}
{Tolman}, R.~C. (1939).
\newblock {Static Solutions of Einstein's Field Equations for Spheres of
  Fluid}.
\newblock \emph{Physical Review} 55, 364--373.
\newblock \doi{10.1103/PhysRev.55.364}
\bibAnnoteFile{Tolman_1939_PhRev}

\bibitem[{Tong et~al.(2022)Tong, Wang, and Wang}]{Tong_2022_apj}
Tong, H., Wang, C., and Wang, S. (2022).
\newblock {Nuclear Matter and Neutron Stars from Relativistic
  Brueckner\textendash{}Hartree\textendash{}Fock Theory}.
\newblock \emph{Astrophys. J.} 930, 137.
\newblock \doi{10.3847/1538-4357/ac65fc}
\bibAnnoteFile{Tong_2022_apj}

\bibitem[{{Vaeth} and {Chanmugam}(1992)}]{Vaeth_1992_A&A}
{Vaeth}, H.~M. and {Chanmugam}, G. (1992).
\newblock {Radial oscillations of neutron stars and strange stars}.
\newblock \emph{Astronomy and Astrophysics} 260, 250--254
\bibAnnoteFile{Vaeth_1992_A&A}

\bibitem[{{Vartanyan} et~al.(2012){Vartanyan}, {Hajyan}, {Grigoryan}, and
  {Sarkisyan}}]{Vartanyan_2012_Ap}
{Vartanyan}, Y.~L., {Hajyan}, G.~S., {Grigoryan}, A.~K., and {Sarkisyan}, T.~R.
  (2012).
\newblock {Stability valley for strange dwarfs}.
\newblock \emph{Astrophys.} 55, 98--109.
\newblock \doi{10.1007/s10511-012-9216-y}
\bibAnnoteFile{Vartanyan_2012_Ap}

\bibitem[{{Vartanyan} et~al.(2014){Vartanyan}, {Hajyan}, {Grigoryan}, and
  {Sarkisyan}}]{Vartanyan_2014_JPhCS}
{Vartanyan}, Y.~L., {Hajyan}, G.~S., {Grigoryan}, A.~K., and {Sarkisyan}, T.~R.
  (2014).
\newblock {Stability of strange dwarfs: a comparison with observations}.
\newblock In \emph{Journal of Physics Conference Series}. vol. 496 of
  \emph{Journal of Physics Conference Series}, 012009.
\newblock \doi{10.1088/1742-6596/496/1/012009}
\bibAnnoteFile{Vartanyan_2014_JPhCS}

\bibitem[{Walecka(1975)}]{Walecka_1975_plb}
Walecka, J.~D. (1975).
\newblock {Equation of State for Neutron Matter at Finite T in a Relativistic
  Mean-Field Theory}.
\newblock \emph{Phys. Lett. B} 59, 109--112.
\newblock \doi{10.1016/0370-2693(75)90678-4}
\bibAnnoteFile{Walecka_1975_plb}

\bibitem[{{Wang} et~al.(2021){Wang}, {Kuerban}, {Geng}, {Xu}, {Zhang}, {Zuo}
  et~al.}]{Wang_2021_PhRvD}
{Wang}, X., {Kuerban}, A., {Geng}, J.-J., {Xu}, F., {Zhang}, X.-L., {Zuo},
  B.-J., et~al. (2021).
\newblock {Tidal deformability of strange quark planets and strange dwarfs}.
\newblock \emph{Phys. Rev. D.} 104, 123028.
\newblock \doi{10.1103/PhysRevD.104.123028}
\bibAnnoteFile{Wang_2021_PhRvD}

\bibitem[{{Wang} et~al.(2000){Wang}, {Dai}, {Lu}, {Wei}, and
  {Huang}}]{Wang_2000_A&A}
{Wang}, X.~Y., {Dai}, Z.~G., {Lu}, T., {Wei}, D.~M., and {Huang}, Y.~F. (2000).
\newblock {A possible model for the supernova/gamma-ray burst connection}.
\newblock \emph{Astron. Astrophys.} 357, 543--547.
\newblock \doi{10.48550/arXiv.astro-ph/9910029}
\bibAnnoteFile{Wang_2000_A&A}

\bibitem[{{Wang} et~al.(2019){Wang}, {Zhou}, {Wang}, and {Liu}}]{Wang_2019_RAA}
{Wang}, Y.-B., {Zhou}, X., {Wang}, N., and {Liu}, X.-W. (2019).
\newblock {The r-mode instability windows of strange stars}.
\newblock \emph{Res. Astron. Astrophys.} 19, 030.
\newblock \doi{10.1088/1674-4527/19/2/30}
\bibAnnoteFile{Wang_2019_RAA}

\bibitem[{{Weber}(2005)}]{Weber_2005_PrPNP}
{Weber}, F. (2005).
\newblock {Strange quark matter and compact stars}.
\newblock \emph{Progress in Particle and Nuclear Physics} 54, 193--288.
\newblock \doi{10.1016/j.ppnp.2004.07.001}
\bibAnnoteFile{Weber_2005_PrPNP}

\bibitem[{Witten(1984)}]{Witten_1984_PhysRevD}
Witten, E. (1984).
\newblock Cosmic separation of phases.
\newblock \emph{Phys. Rev. D.} 30, 272--285.
\newblock \doi{10.1103/PhysRevD.30.272}
\bibAnnoteFile{Witten_1984_PhysRevD}

\bibitem[{{Yip} et~al.(2023){Yip}, {Chi-Kit Cheong}, and {Li}}]{Yip_2023_arXiv}
{Yip}, A. K.~L., {Chi-Kit Cheong}, P., and {Li}, T. G.~F. (2023).
\newblock {Gravitational wave signatures from the phase-transition-induced
  collapse of a magnetized neutron star}.
\newblock \emph{arXiv e-prints} ,
  arXiv:2305.15181\doi{10.48550/arXiv.2305.15181}
\bibAnnoteFile{Yip_2023_arXiv}

\bibitem[{{Zdunik}(2002)}]{Zdunik_2002_A&A}
{Zdunik}, J.~L. (2002).
\newblock {On the minimum radius of strange stars with crust}.
\newblock \emph{Astron. Astrophys.} 394, 641--645.
\newblock \doi{10.1051/0004-6361:20021177}
\bibAnnoteFile{Zdunik_2002_A&A}

\bibitem[{{Zhang} et~al.(2024){Zhang}, {Zou}, {Huang}, {Gao}, {Wang}, {Cui}
  et~al.}]{Zhang_2024_MNRAS}
{Zhang}, X.-L., {Zou}, Z.-C., {Huang}, Y.-F., {Gao}, H.-X., {Wang}, P., {Cui},
  L., et~al. (2024).
\newblock {Gravitational wave emission from close-in strange quark planets
  around strange stars with magnetic interactions}.
\newblock \emph{Mon. Not. R. Astron. Soc.} 531, 3905--3911.
\newblock \doi{10.1093/mnras/stae1400}
\bibAnnoteFile{Zhang_2024_MNRAS}

\bibitem[{{Zhang} et~al.(2018){Zhang}, {Geng}, and {Huang}}]{Zhang_2018_ApJ}
{Zhang}, Y., {Geng}, J.-J., and {Huang}, Y.-F. (2018).
\newblock {Fast Radio Bursts from the Collapse of Strange Star Crusts}.
\newblock \emph{Astrophys. J.} 858, 88.
\newblock \doi{10.3847/1538-4357/aabaee}
\bibAnnoteFile{Zhang_2018_ApJ}

\bibitem[{{Zhao} et~al.(2010){Zhao}, {Cao}, {Luo}, {Sun}, and
  {Zong}}]{Zhao_2010_MPLA}
{Zhao}, A.~M., {Cao}, J., {Luo}, L.-J., {Sun}, W.-M., and {Zong}, H.-S. (2010).
\newblock {The Equation of State of Quasi-Particle Model of Quark-Gluon Plasma
  at Finite Chemical Potential}.
\newblock \emph{Mod. Phys. Lett. A.} 25, 47--54.
\newblock \doi{10.1142/S0217732310031361}
\bibAnnoteFile{Zhao_2010_MPLA}

\bibitem[{{Zong} and {Sun}(2008{\natexlab{a}})}]{Zong_2008_IJMPA}
{Zong}, H.-S. and {Sun}, W.-M. (2008{\natexlab{a}}).
\newblock {a Model Study of the Equation of State of QCD}.
\newblock \emph{Int. J. Mod. Phys. A.} 23, 3591--3612.
\newblock \doi{10.1142/S0217751X08040457}
\bibAnnoteFile{Zong_2008_IJMPA}

\bibitem[{{Zong} and {Sun}(2008{\natexlab{b}})}]{Zong_2008_PhRvD}
{Zong}, H.-S. and {Sun}, W.-M. (2008{\natexlab{b}}).
\newblock {Calculation of the equation of state of QCD at finite chemical and
  zero temperature}.
\newblock \emph{Phys. Rev. D.} 78, 054001.
\newblock \doi{10.1103/PhysRevD.78.054001}
\bibAnnoteFile{Zong_2008_PhRvD}

\bibitem[{{Zou} and {Huang}(2022)}]{Zou_2022_ApJ}
{Zou}, Z.-C. and {Huang}, Y.-F. (2022).
\newblock {Gravitational-wave Emission from a Primordial Black Hole Inspiraling
  inside a Compact Star: A Novel Probe for Dense Matter Equation of State}.
\newblock \emph{Astrophys. J. Lett.} 928, L13.
\newblock \doi{10.3847/2041-8213/ac5ea6}
\bibAnnoteFile{Zou_2022_ApJ}

\bibitem[{{Zou} et~al.(2022){Zou}, {Huang}, and {Zhang}}]{2022Univ....8..442Z}
{Zou}, Z.-C., {Huang}, Y.-F., and {Zhang}, X.-L. (2022).
\newblock {Gravitational Waves from Strange Star Core{\textendash}Crust
  Oscillation}.
\newblock \emph{Universe} 8, 442.
\newblock \doi{10.3390/universe8090442}
\bibAnnoteFile{2022Univ....8..442Z}

\bibitem[{{Zou} et~al.(2021){Zou}, {Zhang}, {Huang}, and {Zhao}}]{Zou_2021_ApJ}
{Zou}, Z.-C., {Zhang}, B.-B., {Huang}, Y.-F., and {Zhao}, X.-H. (2021).
\newblock {Gamma-Ray Burst in a Binary System}.
\newblock \emph{Astrophys. J.} 921, 2.
\newblock \doi{10.3847/1538-4357/ac1b2d}
\bibAnnoteFile{Zou_2021_ApJ}

\end{thebibliography}


\section*{Figure captions}




\end{document}